\documentclass[pra, twocolumn, amsmath,amssymb, superscriptaddress, nofootinbib, floatfix ]{revtex4-1}

\usepackage{multirow}
\usepackage{graphicx}
\usepackage{amsfonts,tikz}  
\usetikzlibrary{matrix}
\usepackage{epstopdf}
\usepackage[version=3]{mhchem}
\usepackage{color}
\usepackage{hyperref}
\usepackage{enumitem}
\usepackage[scr=boondoxo,scrscaled=1.25]{mathalfa}
\usepackage{scalerel}

\hypersetup{%
  pdfpagemode=None, 
  pdfstartpage=1,
  pdfstartview=FitH,
  pdfmenubar=true,
  pdftoolbar=true,
  colorlinks = true,
  linkcolor=blue,
  citecolor=blue,
  urlcolor = blue,
  bookmarksopen=false
}

\newcommand{\Op}[1]{\ensuremath{\boldsymbol{\mathsf{\hat{#1}}}}}

\newcommand{\op}[1]{\ensuremath{\mathsf{\hat{#1}}}}

\newcommand{\Bra}[1]{\ensuremath{\left\langle #1 \right\vert}}
\newcommand{\KetBra}[2]{\Ket{#1}\kern-0.1em\Bra{#2}}
\newcommand{\Ket}[1]{\ensuremath{\left\vert #1 \right\rangle}}

\definecolor{Pantone268}{cmyk}{0.82,1.0,0.0,0.12}
\definecolor{HunterOrange}{cmyk}{0.0,0.55,1.0,0.0}

\begin{document}

%% \title{Manipulation of quantum paths for the control of photoionization dynamics
%% in multi-color polarization-shaped fields}
\title{Perfect control of photoelectron anisotropy for randomly oriented ensembles of
molecules by XUV REMPI and polarization shaping}
\date{\today}

\author{R.~Esteban Goetz}
%\email{egoetz@phys.ksu.edu}
\affiliation{Department of  Physics, Kansas State University, 116 Cardwell Hall, 
1228 N. 17th St. Manhattan, KS 66506-2601 }

\author{Christiane P. Koch}
\affiliation{Theoretische Physik, Universit\"{a}t Kassel, Heinrich-Plett-Str. 40, 
D-34132 Kassel, Germany}
%\email{christiane.koch@uni-kassel.de}

\author{Loren Greenman}
\email{lgreenman@phys.ksu.edu}
\affiliation{Department of Physics, Kansas State University, 116 Cardwell Hall, 
1228 N. 17th St. Manhattan, KS 66506-2601 }

\begin{abstract}
    We report two schemes to generate perfect 
    anisotropy in the photoelectron
    angular distribution of a randomly oriented ensemble of polyatomic molecules. 
    In order to exert full control over the anisotropy of photoelectron
    emission, we exploit interferences between single-photon 
    pathways and a manifold of resonantly-enhanced two-photon %ionization 
    pathways. These are shown to outperform non-sequential $(\omega, 2\omega)$ bichromatic  phase control for the example of  CHFClBr molecules.  
    We are able to
    optimize pulses that yield anisotropic photoelectron emission thanks to a very efficient calculation of
    photoelectron momentum distributions. This is accomplished by combining 
    elements of quantum chemistry, variational scattering theory, and
    time-dependent perturbation theory.
\end{abstract}

\maketitle
%----------------------------------------------------------------------------------------------------------------------------------------

\section{Introduction}
\label{sec:Introduction}

Modern XUV and x-ray sources are increasing in brightness, time
resolution, and phase stability~\cite{allaria2012highly,spezzani2011coherent}, and these advances will lead to the use
of light to probe and control the dynamics of electrons in molecules on their
natural timescales. Additionally, coincidence measurement 
techniques~\cite{brehm1967koinzidenzmessung,baer1991photoelectron,Bodi2009} and
laser alignment~\cite{normand1992laser,Stapelfeldt_2004,horn2006adaptive} are improving the ability to recover the molecular frame
in XUV and x-ray experiments. However, at light sources where it is impractical
to perform coincidence experiments, or in systems of growing complexity where
alignment or analysis of the fragmentation is difficult, complementary methods
are required to obtain sensitive, differential information. The anisotropic
photoelectron distributions induced by breaking parity symmetry are one example
of a possible complementary
technique~\cite{PowisAdvCP08,bowering2001asymmetry,nahon2006determination,harding2005circular,LuxCPC15,yuan2013photoelectron,YIN1995591,douguet2016photoelectronCirc,grum2015photoelectron,douguet2017above,gryzlova2018quantum,yuan2016interference,hu2019coherent,yuan2016photoelectron,Demekhin2018,goetz2019quantum}.
In the photoelectron circular dichroism (PECD)
technique~\cite{LehmannJCP13,JanssenPCCP14,FanoodNatureComm15,FanoodPCCP15,JanssenPhysChemChemPhys2014,PowisAdvCP08,LuxCPC15,beaulieu2018photoexcitation,beaulieu2016universality,kastner2016enantiomeric,miles2017new}, 
chiral molecules are used to break parity symmetry, and the differential photoelectron angular
distribution (PAD) for ionization by left and right circularly polarized
light is measured. This technique has recently been extended to time-resolved
studies~\cite{comby2016relaxation,beaulieu2016probing}, illustrating its promise as a probe of dynamics. However, in
the XUV and x-ray regimes, the techniques for generation and control % lg-24/ of stable
of highly coherent circularly polarized
light sources are limited. 
Also, differential techniques complementary to coincidence measurements that can
probe the structure and dynamics in achiral molecules are also desired.

An alternative to PECD is the control of anisotropy 
in a single PAD using multiphoton
pathways, where the light fields %can be 
are used to break parity symmetry~\cite{shapiro2012quantum}. Such
studies generally focus on using two-color pulses to manipulate the phase 
between two quantum pathways.
Two-pathway coherent control of the PAD in non-sequential bichromatic $(\omega, 2\omega)$ photoionization  
has been reported in atomic systems~\cite{Muller1990,YinPRL92,SchumacherPRL94,gryzlova2018quantum,douguet2016photoelectronCirc,grum2015photoelectron}, which are invariant under rotation operations. 
In particular, a high degree of left-right 
asymmetry $(\approx 100\%)$ has been reported in theoretical studies of atomic
hydrogen~\cite{GrumPRA2015} and neon~\cite{gryzlova2018quantum} by 
interfering a single-photon ionization channel and one resonantly enhanced two-photon ionization pathway. 

In molecular systems, on the other hand, single and multi-photon ionization processes are 
highly spherically asymmetric~\cite{underwood2000time,suzuki2005mapping,son2009multielectron,hockett2015general}. Consequently, frame-rotation effects 
can be observed in bichromatic coherent control of asymmetries in the molecular PAD~\cite{YIN1995591,yuan2016photoelectron,ArtemyevJCP5}. 
In this context, a high degree of anisotropy $(100\%)$ has been measured in the phase-controlled 
bichromatic ionization of  aligned molecular \ce{NO}~\cite{YIN1995591}, and it
has been 
calculated with nuclear motion for aligned ~\ce{H^+_2}~\cite{yuan2013photoelectron,yuan2016interference}. 
The sensitivity of the PAD to the field helicity 
may also be exploited for the purpose of controlling the asymmetry in the PAD. For instance, a high-degree of
asymmetry  from (pre-aligned)  
single-electron $\ce{H^+_2}$ was reported in phase-controlled bichromatic ionization
using co- and counter-rotating field polarization components of attosecond UV
fields~\cite{yuan2016photoelectron}.

While the coherent control calculations in
Refs.~\cite{yuan2013photoelectron,yuan2016interference,yuan2016photoelectron} assume molecular alignment, the experimental conditions might be such that the  initial orientation 
of the target cannot be unambiguously defined. Consequently, an equi-probable orientation
distribution is often assumed by integrating over all possible molecular
orientations with a homogeneous probability distribution~\cite{RitchiePRA76,ChandraPRA87,ChandraJPB87}.
Without laser alignment techniques, the efficiency of two-color control of
anisotropy may be obscured by orientation averaging, or it may be completely suppressed.   
Although the orientation averaging approach has become the gold standard for theoretical  
studies on chiro-optical discrimination in rotationally
isotropic
media~\cite{GoetzJCP17,goetz2019quantum,beaulieu2018multiphoton,LeinPRA2014,Demekhin2018},
the question of whether the anisotropy in the PAD 
persists after the orientation averaging in 
linearly polarized bichromatic ionization remains yet to be answered. 

Polarization-shaped pulses, wherein the instantaneous polarization~\cite{misawa2016applications}
or helicity
changes dynamically over time~\cite{brixner2001femtosecond,brixner2002generation,brixner2004quantum,plewicki2006phase,selle2008generation,ninck2007programmable}  
offer another degree of freedom for control~\cite{brixner2004quantum,hockett2015coherent,kerbstadt2017control}. 
However, the efficiency of shaping the polarization of the driving field 
in the specific context of resonantly-enhanced multi-photon ionization (REMPI)
to achieve perfect anisotropy in a randomly ensemble of molecules is, to the best of our knowledge, 
not known.  In particular, whether 
non-sequential bichromatic $(\omega, 2\omega)$ phase control~\cite{shapiro2012quantum} 
or sequential wave packet evolution-based pump-probe~\cite{TannorJCP85} schemes 
suffice to achieve perfect anisotropy in randomly oriented molecules or whether a more general control scheme
based on coherent control~\cite{misawa2016applications} is needed remains an open question. 

In order to answer these questions,
we first show that the anisotropy 
in linearly polarized bichromatic ionization does persist after orientation
averaging. As a second step, we identify    
the limitations of this approach to
achieve perfect anisotropy in a 
randomly oriented ensemble of CHFClBr molecules.
We then demonstrate how to achieve perfect anisotropy by exploiting quantum 
pathway interferences between
single-photon ionization pathways and a manifold of REMPI
paths driven by linearly polarized multi-color
fields. 

Additionally, we investigate the influence of the polarization state (linear, circular left, right) of the driving field, 
and extend our analysis to the case of polarization shaped pulses. We optimize
the time-dependence of the polarization state 
by combining fields with simultaneous counter-rotating components~\cite{hockett2015coherent,kerbstadt2017control}, 
i.e., by combining multicolor fields circularly polarized along left and right
polarization directions. We show that
quantum interferences driven by polarization-shaped fields 
results in perfect anisotropy in the orientation-averaged PAD. 
We find that the individual contribution of each
circularly polarized component induces very modest asymmetry, whereas a
combination of components leads to a much larger effect.

We are able to find the optimal REMPI pathways using quantum optimal control
theory~\cite{Glaser2015}. 
This requires a method for calculating the photoionization
dynamics of molecules that can be repeated for a number of different laser
pulses efficiently. We use a combination of quantum chemistry to describe the
bound states, variational scattering theory to calculate dipole matrix elements
between bound states and photoionized states, and time-dependent perturbation
theory to describe the dynamics. We have implemented this technique in
Ref.~\cite{goetz2019quantum} and  have extended it here to pulses with
arbitrary polarization state.

This work is organized as follows. In Section~\ref{sec:TheoreticalFramework}, we
present the details of the derivation of the orientation-averaged PAD. 
In Section~\ref{subsec:CoherentControlBichro}, we construct a control scheme based on 
multiple REMPI pathways and compare  
its performance in maximizing the anisotropy of the PADs of a 
randomly oriented ensemble of $\ce{CHBrClF}$ molecules
against that of the two-color coherent control driven by bichromatic $(\omega,
2\omega)$ pulses. 
Finally, we extend our findings to the case of polarization-shaped
pulses in Sec.~\ref{subsec:polshaping} and Sec.~\ref{sec:Conclusion} concludes. 

\section{Theoretical Framework}
\label{sec:TheoreticalFramework}

\subsection{Laboratory-frame orientation averaged PAD}
\label{subsec:Manybodytreatment}

We first detail our methodology to calculate the orientation averaged photoelectron 
momentum distribution in the laboratory frame of reference, which is formulated in the strict electric dipole approximation.
In what follows primed and unprimed bold symbols are used to define vector
 quantities 
in the fixed laboratory $(\mathcal{R}^\prime)$  and molecular $(\mathcal{R})$
frames of reference, respectively, with
$\mathcal{R}$ being rotated relative to $\mathcal{R}^\prime$ by Euler
angles $\gamma_{\mathcal{R}}=(\alpha,\beta,\gamma)$~\cite{edmonds2016angular}. Neglecting relativistic effects and
assuming fixed nuclei during the interaction,                                 
the  Schr\"odinger equation for the many electron system in $\mathcal{R}$ reads 
\begin{eqnarray}
 \label{eq:SCHCIS0}
  i\dfrac{\partial}{\partial\,t}|\Psi^N(t;\gamma_{\mathcal{R}})\rangle  &=& \Big[\op{\mathcal{H}}_0 - \text{\textbf{E}}(t;\gamma_{\mathcal{R}})\cdot\Op{r}\Big]|\Psi^N(t;\gamma_{\mathcal{R}})\rangle\,,\quad
 \end{eqnarray}
where $\op{\mathcal{H}}_0 = \op{H}_0 + \op{H}_1$ refers to field-free Hamiltonian,
with $\op{H}_0$ and $\op{H}_1$ the mean-field Fock operator and the
 residual Coulomb interaction~\cite{GreenmanPRA2010}, respectively. 
  Finally, $\text{\textbf{E}}(t;\gamma_{\mathcal{R}})$
is the electric field in $\mathcal{R}$.
The polarization components of the driven field are known in the laboratory frame. It can thus be defined 
in terms of the (fixed) spherical unit vectors,
 $\text{\textbf{e}}^\prime_{\mu_0}$,  with $\mu_0=\pm 1,0$~\cite{edmonds2016angular}, relative to $\mathcal{R}^\prime$, namely 
 \begin{subequations} 
 \begin{eqnarray}
   \label{eq:Efield.def}
   \text{\textbf{E}}^\prime(t)=\sum_{\mu_0=0,\pm 1}\mathscr{E}^{\prime}_{\mu_0}(t)\,\text{\textbf{e}}^{*\prime}_{\mu_0},
 \end{eqnarray}
 where $(*)$ denotes the complex conjugation, 
   and where $\mathscr{E}^\prime_{\mu_0}(t)$ are the polarization unit components of the field in
   $\mathcal{R}^\prime$. The cartesian components of the spherical unit vectors 
   are defined in the usual manner and given in Eq.~\eqref{eq:cart_to_sph} in Appendix~\ref{sec:appendix1}. Upon projection  
   of $\text{\textbf{e}}^{*\prime}_{\mu_0}$ 
   into $\mathcal{R}$, as detailed in Appendix~\ref{sec:appendix1}, the  
   molecular-frame orientation-dependent dipole interaction reads
 \begin{eqnarray}
   \label{eq:dipol_mol_fram}
   \text{\textbf{E}}(t;\gamma_{\mathcal{R}})\cdot\Op{r}=\sum_{\mu_0}(-1)^{\mu_0}\mathscr{E}^\prime_{\mu_0}(t)\sum_{\mu}\mathcal{D}^{(1)}_{\mu, -\mu_0}(\gamma_{\mathcal{R}})\,\Op{r}_{\mu}\,,\quad\quad 
 \end{eqnarray}
  \end{subequations}
  where $\mathcal{D}^{(1)}_{\mu,\mu_{0}}(\gamma_{\mathcal{R}})$ are the
 elements of the Wigner rotation matrix~\cite{edmonds2016angular,rose1957elementary}. 
 
 Accounting for one-particle one-hole excitations only, the many-body
  wave function is described by the TDCIS ansatz~\cite{KlamrothPRB03}
\begin{eqnarray}
\label{eq:cis_expansion} 
  |\Psi^N(t;\gamma_{\mathcal{R}})\rangle &=&
  \alpha_0(t;\gamma_{\mathcal{R}})\,e^{-i\varepsilon_{o}t}\,|\Phi_0\rangle\nonumber \\[0.2cm] 
   && +\sum_{i,a}\alpha^{a}_{i}(t;\gamma_{\mathcal{R}})\,e^{-i\varepsilon_i^a t}\,|\Phi^{a}_{i}\rangle\\\nonumber
    &&+\sum_{i}\int \mathrm{d}\boldsymbol{k}\,\alpha^{\boldsymbol{k}}_{i}(t;\gamma_{\mathcal{R}})\,e^{-i\varepsilon_i^k t}\,|\Phi^{\boldsymbol{k}}_{i}\rangle\,,
\end{eqnarray}
where $\alpha_0(t;\gamma_{\mathcal{R}})$, $\alpha^a_i(t;\gamma_{\mathcal{R}})$ and $\alpha^{\boldsymbol{k}}_{i}(t;\gamma_{\mathcal{R}})$ are time-dependent coefficients, and 
$|\Phi_0\rangle$ 
refers to the Hartree-Fock ground state.
$|\Phi^{a}_{i}\rangle = \Op{c}^\dagger_a\Op{c}_i|\Phi_0\rangle$  
 describes the
one-particle one-hole excitation from an initially occupied
orbital $\varphi_i$ to an initially unoccupied Hartree-Fock orbital $\varphi_a$ with orbital energy $\epsilon_a$, whereas $|\Phi^{\boldsymbol{k}}_{i}\rangle$ 
describes the excitation to scattering continuum state 
$\varphi^-_{\boldsymbol{k}}$ with energy $\vert\boldsymbol{k}\vert^2/2$, respectively. 
We denote the Fock energy of a single determinant as $\varepsilon$, e.g.,
$\varepsilon_0=\sum_i \epsilon_i$.
For the calculations presented here, we further restrict the configuration space
in Eq.~\eqref{eq:cis_expansion} to excitations from the highest occupied molecular orbital (HOMO, labeled $i_0$) only.
The Hartree-Fock orbitals  
were obtained using
the \texttt{MOLPRO}~\cite{werner2012molpro,werner2012molprowires} program
package at the \texttt{aug-cc-pVDZ} basis set~\cite{augccpvdz} level.

Neglecting the residual Coulomb interaction, the coupled equations for the expansion coefficients read, 
\begin{widetext}
\begin{subequations}
\label{eq:coupled.set}
\begin{eqnarray}
  \label{eq:a0.diff}
  \dot{\alpha_0}(t;\mathcal{\gamma_{R}}) &=&
    i\sum_{\mu_0,\mu}(-1)^{\mu_0}\mathcal{D}^{(1)}_{\mu,-\mu_0}(\gamma_{\mathcal{R}})\mathscr{E}^\prime_{\mu_0}(t) \\
  &&\times
  \Big[\sum_{i} \left(\mathbf{r}_{i,i}\cdot\mathbf{\textbf{e}}_{\mu}\right)\alpha_0(t;\gamma_{\mathcal{R}})
                    +\sum_{i,a}\left(\mathbf{r}_{i,a}\cdot\mathbf{\textbf{e}}_{\mu}\right) e^{i(\epsilon_i-\epsilon_a)t}\,\alpha^a_i(t;\gamma_{\mathcal{R}})
                    +\sum_{i}\int\!\! \mathrm{d}\boldsymbol{k}\,\left(\mathbf{r}_{i,\boldsymbol{k}}\cdot\mathbf{\textbf{e}}_{\mu}\right)
                    e^{i(\epsilon_i-\epsilon_k)t}\,\alpha^{\boldsymbol{k}}_i(t;\gamma_{\mathcal{R}})\Big],\nonumber\\[0.3cm]
  \label{eq:aia.diff}
  \dot{\alpha}^a_i(t;\gamma_{\mathcal{R}}) &=&  i\sum_{\mu_0,\mu}(-1)^{\mu_0}\mathcal{D}^{(1)}_{\mu,-\mu_0}(\gamma_{\mathcal{R}})\mathscr{E}^\prime_{\mu_0}(t)\\
  &&\times\Big[ \left(\mathbf{r}_{a,i}\cdot\textbf{e}_\mu\right) e^{i(\epsilon_a-\epsilon_i)t} \alpha_0(t;\gamma_{\mathcal{R}})
    +\sum_{b\neq{}a} \left(\mathbf{r}_{{a,b}}\cdot\textbf{e}_{\mu}\right) e^{i(\epsilon_a-\epsilon_b)t}\,\alpha^b_i(t;\gamma_{\mathcal{R}})
    -\sum_{j\ne i}\left(\mathbf{r}_{i,j}\cdot\textbf{e}_{\mu}\right) e^{i(\epsilon_j-\epsilon_i)t}\,\alpha^a_j(t;\gamma_{\mathcal{R}})
    \nonumber\\
    &&\quad\quad + \Big((\sum_j \mathbf{r}_{j,j} - \mathbf{r}_{i,i} +
    \mathbf{r}_{a,a})\cdot\textbf{e}_{\mu}\Big)\, \alpha_i^a(t;\gamma_{\mathcal{R}}) 
    +\int{}\!\mathrm{d}\boldsymbol{k} \left(\mathbf{r}_{a,\boldsymbol{k}}\cdot\textbf{e}_{\mu}\right) e^{i(\epsilon_a-\epsilon_k)t}\,\alpha^{\boldsymbol{k}}_i(t;\gamma_{\mathcal{R}})
    \Big]\,,\nonumber \\[0.3cm]
  \label{eq:aik.diff}
  \dot{\alpha}^{\boldsymbol{k}}_i(t;\gamma_{\mathcal{R}}) &=&  i\sum_{\mu_0,\mu}(-1)^{\mu_0}\mathcal{D}^{(1)}_{\mu,-\mu_0}(\gamma_{\mathcal{R}})\mathscr{E}^\prime_{\mu_0}(t)\\
  &&\times\Big[ \left(\mathbf{r}_{\boldsymbol{k},i}\cdot\textbf{e}_{\mu}\right) e^{i(\epsilon_{\boldsymbol{k}}-\epsilon_i)t} \alpha_0(t;\gamma_{\mathcal{R}})
    +\sum_{b} \left(\mathbf{r}_{{\boldsymbol{k},b}}\cdot\textbf{e}_{\mu}\right) e^{-i(\epsilon_b-\epsilon_{\boldsymbol{k}})t}\,\alpha^b_i(t;\gamma_{\mathcal{R}})
   -\sum_{j}\left(\mathbf{r}_{{i,j}}\cdot\textbf{e}_{\mu}\right) e^{i(\epsilon_j-\epsilon_i)t}\,\alpha^{\boldsymbol{k}}_j(t;\gamma_{\mathcal{R}}) 
    \nonumber\\
    &&\quad\quad + \Big((\sum_j\mathbf{r}_{j,j} - \mathbf{r}_{i,i} +
    \mathbf{r}_{\boldsymbol{k},\boldsymbol{k}})\cdot\textbf{e}_{\mu}\Big)\,\alpha_{i}^{\boldsymbol{k}}(t;\gamma_\mathcal{R})
  +\int_{\boldsymbol{k}^\prime\ne\boldsymbol{k}}\!\mathrm{d}\boldsymbol{k}^\prime \left(\mathbf{r}_{{\boldsymbol{k}\boldsymbol{k}^\prime}}\cdot\textbf{e}_{\mu}\right) e^{i(\epsilon_{k}-\epsilon_{k^\prime})t}\,\alpha^{\boldsymbol{k}^\prime}_i(t;\gamma_{\mathcal{R}})\Big]\,.\nonumber 
  %&&-i \sum_{j,b}\langle\Phi^a_i|\Op{H}_1|\Phi^b_j\rangle\alpha^b_j(t;\gamma_{\mathcal{R}})e^{-(\epsilon^b_j - \epsilon^a_i)t}\,.\nonumber
\end{eqnarray}
\end{subequations}
\end{widetext}
where $\boldsymbol{r}_{p,q}\cdot\textbf{e}_{\mu}=\langle\varphi_p|\op{r}_\mu|\varphi_q\rangle$. The  coefficients $\alpha^{\boldsymbol{k}}_{i}(t;\gamma_{\mathcal{R}})$
describe the transition amplitude from an initially occupied 
orbital $i$ to a continuum state with energy $\varepsilon_k =
\varepsilon_0-\epsilon_i+|\boldsymbol{k}|^2/2$ in the direction
$\boldsymbol{k}/|\boldsymbol{k}|$ with respect to the molecular frame of reference,
$\mathcal{R}$. 
Since this state is not an eigenstate of the Fock
operator~\cite{lucchese1986applications}, it is an assumption that it can be written as
such in Eq.~\eqref{eq:aik.diff}. 
Similarly,  
$\alpha^{\boldsymbol{k}^\prime}_{i}(t;\gamma_{\mathcal{R}})$ describe this transition
in the laboratory frame, $\mathcal{R}^\prime$.

To model an ensemble of randomly oriented 
molecules, we average                                                           
over all Euler angles $\gamma_{\mathcal{R}}$.  
The orientation-averaged photoelectron momentum distribution is 
obtained upon integration over $\gamma_{\mathcal R}$ and 
incoherent summation over the initially occupied contributing orbitals $i$ 
in the Hartree-Fock ground state, 
\begin{eqnarray}
 \label{eq:ExactForm}
 \dfrac{d^2\sigma}{d\epsilon_k\,d\Omega_{\boldsymbol{k}^\prime}} &=&  
  \sum_{i\in\{\text{occ}\}}\int
    |\alpha_i^{\boldsymbol{k}^\prime}(t;\gamma_{\mathcal{R}})|^2\, \mathrm{d}^3\gamma_{\mathcal{R}}\,, 
\end{eqnarray} 
for $t\rightarrow\infty$ and
with $\boldsymbol{k}^\prime$ denoting the
momentum measured in the laboratory frame.  
We illustrate how to transform the TDCIS coefficients to the
laboratory frame in Sec.~\ref{subsec:Variationalscatteringstates}.

\subsection{Electron dynamics: Time-dependent perturbative treatment}
\label{subsec:perturbationtreatment}

The photoionization process is 
captured by the coefficients $\alpha^{\boldsymbol{k}^\prime}_i(t;\gamma_{\mathcal{R}})$ and
requires an accurate description of the scattering components of the
wave function in Eq.~\eqref{eq:cis_expansion}. For a many-electron system with no symmetry, exact numerical
simulation of the electron dynamics represents a formidable computational challenge
with prohibitive computational cost. We circumvent this by solving Eq.~\eqref{eq:coupled.set} perturbatively. 
A second-order                                        
approximation is suitable to manipulate quantum interferences
between  conventional opposite-parity pathways to control the anisotropy of photoelectron 
emission~\cite{douguet2016photoelectronCirc,grum2015photoelectron,douguet2017above,gryzlova2018quantum}.
It can also  describe the necessary dynamics of
same-parity (two-photon) pathways~\cite{goetz2019quantum}.
Equation~\eqref{eq:ExactForm}  simplifies to 
\begin{eqnarray}
 \label{eq:SecondOrderApprox}
 \dfrac{d^2\sigma}{d\epsilon_k\,d\Omega_{\boldsymbol{k}^\prime}} &\approx&  \int
    \big|\alpha^{\boldsymbol{k}^\prime\,(1)}_{i_0}(t;\gamma_{\mathcal{R}})
    +\alpha^{\boldsymbol{k}^\prime\,(2)}_{i_0}(t;\gamma_{\mathcal{R}})  \big|^2
    \mathrm{d}^3\gamma_{\mathcal{R}}\,,\quad
\end{eqnarray} 
for $t\to\infty$ and with $\alpha^{\boldsymbol{k}^\prime\,(1,2)}_{i_0}(t;\gamma_{\mathcal{R}})$, 
the first, resp. second, order correction. 
The differential cross section in Eq.~\eqref{eq:SecondOrderApprox} can be written in terms of the
associated Legendre polynomials $P^M_L(\cdot)$,
\begin{eqnarray}
    \label{eq:betaparameters}
    \dfrac{d^2\sigma}{d\epsilon_k\,d\Omega_{\boldsymbol{k}^\prime}} = \sum_{L,M}
    \beta_{L,M}(\epsilon_k)\,P^M_L(\cos\theta_{\boldsymbol{k}^\prime})\,e^{iM\phi_{\boldsymbol{k}^\prime}}\,.
\end{eqnarray}

Following Ref.~\cite{goetz2019quantum}, we write the photoelectron momentum distribution defined in Eq.~\eqref{eq:SecondOrderApprox} 
in terms of the individual contributions from one- and two-photon ionization processes and
their interference,
\begin{eqnarray}
 \label{eq:SecondOrderApprox_split}
 \dfrac{d^2\sigma}{d\epsilon_k\,d\Omega_{\boldsymbol{k}^\prime}} &=&                                                                                                                                                                                        
 \dfrac{d^2\sigma^{1ph}}{d\epsilon_k\,d\Omega_{\boldsymbol{k}^\prime}}+
 \dfrac{d^2\sigma^{2ph}}{d\epsilon_k\,d\Omega_{\boldsymbol{k}^\prime}}+
 \dfrac{d^2\sigma^{int}}{d\epsilon_k\,d\Omega_{\boldsymbol{k}^\prime}}\,.
\end{eqnarray} 
The contribution from one- and two-photon processes defined by the first two
terms in the rhs. in Eq.~\eqref{eq:SecondOrderApprox_split} reads (for $n=1,2$)
\begin{eqnarray}
  \label{eq:1phcontr}
  \dfrac{d^2\sigma^{nph}}{d\epsilon_k\,d\Omega_{\boldsymbol{k}^\prime}}&=&
    \int\alpha^{(n)\boldsymbol{k}^\prime}_{i_0}(t;\gamma_{\mathcal{R}})\alpha^{*(n)\boldsymbol{k}^\prime}_{i_0}(t;\gamma_{\mathcal{R}})\,\mathrm{d}^3\gamma_{\mathcal{R}}\\         
&=&
\sum_{L,M}\beta^{nph}_{L,M}(\epsilon_k)\,P^M_L(\cos\theta_{\boldsymbol{k}^\prime})\,e^{iM\phi_{\boldsymbol{k}^\prime}}\,.\quad \nonumber
\end{eqnarray}
The expansion coefficients $\beta^{1ph(2ph)}_{L,M}(\epsilon_k)$  correspond to the
orientation-averaged anisotropy
parameters~\cite{reid2003photoelectron} associated with the first (second) order correction. 
Defining the complex-valued term, 
\begin{eqnarray}
  \label{eq:beta.int.def}
  \beta^{int}_{L,M} &=&          
\int
    \alpha^{(1)\boldsymbol{k}^\prime}_{i_0}(\gamma_{\mathcal{R}})\alpha^{*(2)\boldsymbol{k}^\prime}_{i_0}(\gamma_{\mathcal{R}})\,\mathrm{d}^3\gamma_{\mathcal{R}}\,, 
\end{eqnarray}
 the contribution from the interfering pathways to the photoelectron spectrum reads   
\begin{eqnarray}
  \label{eq:12phcontr}
  \dfrac{d^2\sigma^{int}}{d\epsilon_k\,d\Omega_{\boldsymbol{k}^\prime}}&=&
    \int\Big(
    \alpha^{(1)\boldsymbol{k}^\prime}_{i_0}(t;\gamma_{\mathcal{R}})\alpha^{*(2)\boldsymbol{k}^\prime}_{i_0}(t;\gamma_{\mathcal{R}})
    + c.c.\Big)\,\mathrm{d}^3\gamma_{\mathcal{R}}\nonumber\\                
&=&\sum_{L,M}\Big(\beta^{int}_{L,M}(\epsilon_k)\,e^{iM\phi_{\boldsymbol{k}^\prime}} + c.c.\Big)\, P^M_L(\cos\theta_{\boldsymbol{k}^\prime})\,.\nonumber\\
\end{eqnarray}
%or, equivalently,
%\begin{eqnarray}
% \label{eq:12phcontr.final}
% \dfrac{d^2\sigma^{int}}{d\epsilon_k\,d\Omega_{\boldsymbol{k}^\prime}}&=& 2\sum_{L,M}\Big[\text{Re}\big[\beta^{int}_{L,M}(\epsilon_k)\big]\cos(M\phi_{\boldsymbol{k}^\prime})\\          
%&& -\, \text{Im}\big[\beta^{int}_{L,M}(\epsilon_k)\big]\sin(M\phi_{\boldsymbol{k}^\prime})\Big]\,P^M_L(\cos\theta_{\boldsymbol{k}^\prime})\nonumber\,,
%\end{eqnarray}
First ($\alpha^{\boldsymbol{k}^\prime\,(1)}_{i_0}$) and
second-order ($\alpha^{\boldsymbol{k}^\prime\,(2)}_{i_0}$)  terms describe 
direct single-photon photoionization from $\varphi_{i_0}$ to $\varphi^-_{\boldsymbol{k}^\prime}$, and resonant two-photon photoionization from $\varphi_{i_0}$ to $\varphi^-_{\boldsymbol{k}^\prime}$ 
via different unoccupied orbitals $\varphi_a$, respectively.

\subsection{Variational scattering states}
\label{subsec:Variationalscatteringstates}

The scattering states required for evaluation of the dipole matrix elements are obtained from variational scattering theory~\cite{GianturcoJCP94,NatalenseJCP99,Greenman2017variational}.
Assuming no relaxation of the contributing orbitals, the   
total many-body wave function $\Phi^{\boldsymbol{k}}_{i}(\boldsymbol{r}_1,\dots \boldsymbol{r}_N)$ can be
defined,
for any $i\in\{\text{occ}\}$,  as an antisymmetrized product, 
\begin{subequations}
\begin{eqnarray}
  \Phi^{\boldsymbol{k}}_{i}(\boldsymbol{r}_1,\dots \boldsymbol{r}_N ) &=& 
\mathcal{A}_{N}\big[\varphi^-_{\boldsymbol{k}}(\boldsymbol{r}_N);\Phi_{i}(\boldsymbol{r}_{1},\dots  \boldsymbol{r}_{N-1})\big]\,,\quad\quad
\end{eqnarray}
  where $\varphi^-_{\boldsymbol{k}}(\boldsymbol{r}_N) $ corresponds to the (molecular-frame) scattering
component  of
the wave function and  $\Phi_{i}( \boldsymbol{r}_{1},\dots\boldsymbol{r}_{N-1})$ the remaining $N-1$ electron final
state after ionization. 
We obtain $\varphi^{-}_{\boldsymbol{k}}(\mathbf{r})$  
by solving the scattering problem 
\begin{eqnarray}
  \label{eq:epolyscat.supp}
  \left[-\dfrac{\nabla^2}{2} - \dfrac{1}{r} + \Op{V} - \dfrac{k^2}{2} \right]\varphi^{-}_{\boldsymbol{k}}(\mathbf{r})   &=&0 \,,
\end{eqnarray}
\end{subequations}
with scattering boundary conditions~\cite{baertschy2001accurate,miller1987new} for the outgoing wave $\varphi^{-}_{\boldsymbol{k}}(\mathbf{r})$ at
large distances $\boldsymbol{r}\rightarrow \infty$,  
and where $\Op{V}(\mathbf{r})$ describes the short-range part of the electron-ion
interaction. Equation~\eqref{eq:epolyscat.supp} and its dipole
matrix elements  are computed using
a locally modified version of the \texttt{ePolyScat} program 
package~\cite{GianturcoJCP94,NatalenseJCP99,Greenman2017variational}. 
The bound unoccupied Hartree-Fock orbitals that are kept in the time-dependent
perturbation expansion are chosen to be those that are orthogonal to the 
scattering orbitals. In this manner, Gaussian orbitals that attempt
to represent continuum states are discarded. 
In the molecular frame, the direction of photoelectron emission is obtained by
expanding the scattering wave function 
into spherical harmonics, 
\begin{subequations}
\begin{eqnarray}
  \label{eq:epolyexpansion}
  \varphi^-_{\boldsymbol{k}}(\boldsymbol{r}) &=&\sum_{\ell,m}\varphi^-_{k,\ell,m}(\boldsymbol{r})\,Y^{\ell *}_m(\theta_{\boldsymbol{k}},\phi_{\boldsymbol{k}})\, ,
\end{eqnarray}
where $\theta_{\boldsymbol{k}}$ and $\phi_{\boldsymbol{k}}$ correspond to the
polar and azimuthal angles of photoelectron emission 
in the molecular frame. In the laboratory frame, this direction is 
  defined by the angles $(\theta_{\boldsymbol{k}^\prime},\phi_{\boldsymbol{k}^\prime})$, 
which is obtained by projecting Eq.~\eqref{eq:epolyexpansion} into the laboratory frame.
In this frame, the scattering states take the form,
\begin{eqnarray}
  \label{eq:epolyexpansion_labframe}
  \varphi^-_{\boldsymbol{k}^\prime}(\boldsymbol{r}) &=&\sum_{\ell,m,m^\prime}\varphi^-_{k,\ell,m}(\boldsymbol{r})\,\mathcal{D}^{(\ell)\dagger}_{m, m^\prime}(\gamma_{\mathcal{R}})\,Y^{\ell *}_{m^\prime}(\theta_{\boldsymbol{k}^\prime},\phi_{\boldsymbol{k}^\prime})\,.\quad\quad\,
\end{eqnarray}
\end{subequations}
Applying first-order time-dependent perturbation theory to
Eq.~\eqref{eq:a0.diff} and evaluating the individual matrix elements of
Eq.~\eqref{eq:epolyexpansion} %, %the result becomes,
results in 
  \begin{eqnarray}
  \label{eq:alpha_ik_first.final}
  \alpha^{\boldsymbol{k}^\prime(1)}_{i_0}(t;\gamma_{\mathcal{R}}) &=&
  i\sum_{\mu_0,\mu}(-1)^{\mu_0}\sum_{\ell,m, m^\prime}\mathcal{D}^{(1)}_{\mu,-\mu_0}(\gamma_{\mathcal{R}})\mathcal{D}^{(\ell)\dagger}_{m^\prime,m}(\gamma_{\mathcal{R}})\nonumber \\
  &&
  \quad\times 
      (\mathbf{r}_{k,\ell,m;i_0}\cdot\textbf{e}_{\mu})\,
      Y^{\ell}_{m^\prime}(\theta_{\boldsymbol{k}^\prime},\phi_{\boldsymbol{k}^\prime}) 
      \nonumber\\
      &&\quad\times\int_{-\infty}^{t}e^{i(\epsilon_k -\epsilon_{i_0})}\mathscr{E}^\prime_{\mu_0}(t^\prime)\,\mathrm{d}t^\prime\,. 
\end{eqnarray}
The dipole matrix element $\mathbf{r}_{k,\ell,m;i_0}=\langle\varphi^{-}_{k,\ell,m}|\Op{r}|\varphi_{i_0}\rangle$ now displays indices for 
the partial wave quantum numbers $\ell$ and $m$ of the continuum orbital, and the ionized
orbital $\varphi_{i_0}$.
Similarly, the expression for the  second order correction of the scattering
component along the direction of photoelectron emission 
$(\theta_{\boldsymbol{k}^\prime},\phi_{\boldsymbol{k}^\prime})$ %direction 
relative to the laboratory frame becomes
\begin{widetext}
 \begin{eqnarray}
  \label{eq:alpha_ik_sec.preliminary}
  \alpha^{\boldsymbol{k}^\prime(2)}_{i_0}(t;\gamma_{\mathcal{R}}) &=&
  -\sum_{\mu_0,\nu_0}(-1)^{\mu_0+\nu_0}\sum_{\mu,\nu}\mathcal{D}^{(1)}_{\mu,-\mu_0}(\gamma_{\mathcal{R}})\mathcal{D}^{(1)}_{\nu,-\nu_0}(\gamma_{\mathcal{R}})
  \sum_{\ell, m, m^\prime}\mathcal{D}^{(\ell)\dagger}_{m^\prime,m}(\gamma_{\mathcal{R}})Y^{\ell}_{m^\prime}(\theta_{\boldsymbol{k}^\prime},\phi_{\boldsymbol{k}^\prime})\nonumber\\
      &&\quad\times\Big[ \left(\mathbf{r}_{k,\ell,m;i_0}\cdot\textbf{e}_{\mu}\right) \sum_i \left(\mathbf{r}_{i,i}\cdot\textbf{e}_{\nu}\right) \,
      \int^t_{-\infty}e^{-i(\epsilon_{i_0}-\epsilon_k)t^\prime}
        \mathscr{E}^{\prime}_{\mu_0}(t^\prime)\,
      \int_{-\infty}^{t^\prime} \mathscr{E}^{\prime}_{\nu_0}(t^{\prime\prime})
        \,\mathrm{d}t^{\prime\prime}\,\mathrm{d}t^\prime\nonumber\\ %[0.3cm]
      &&\quad\quad +\sum_{b} \left(\mathbf{r}_{k,\ell,m;b}\cdot\textbf{e}_{\mu}\right)
      \left(\mathbf{r}_{b,i_0}\cdot\textbf{e}_{\nu}\right)
      \int^t_{-\infty}e^{-i(\epsilon_{b}-\epsilon_k)t^\prime}\mathscr{E}^{\prime}_{\mu_0}(t^\prime)\,
      \int_{-\infty}^{t^\prime}e^{-i(\epsilon_{i_0}-\epsilon_b)}\mathscr{E}^{\prime}_{\nu_0}(t^{\prime\prime})\,dt^{\prime\prime}\,dt^\prime\Big]\,.  % \\ %[0.3cm]
 \end{eqnarray}
\end{widetext}
In Eq.~\eqref{eq:alpha_ik_sec.preliminary}, we further assume that the last two terms
corresponding to Eq.~\eqref{eq:aik.diff} can be neglected,  which is justified
by the absence of IR and high-energy
XUV photon energies --that are required to make the corresponding time integrals non-zero--
in all pulses considered here.

\subsection{Laboratory-frame orientation-averaged anisotropy parameters}
\label{subsec:anisotropyparameters}

The laboratory-frame orientation-averaged anisotropy parameters 
associated with one- and two-photon ionization pathways and their interference
defined in Eqs.~\eqref{eq:1phcontr} and~\eqref{eq:beta.int.def} can be obtained 
using the expressions defined in
Eqs.~\eqref{eq:alpha_ik_first.final} and ~\eqref{eq:alpha_ik_sec.preliminary}. 
Derivation of the laboratory-frame anisotropy parameters involves cumbersome but straightforward angular momentum algebra. We give explicit details
of the derivations in Appendix~\ref{sec:appendix2}. Here, we provide only their ellipticity dependence in the view of listing a few selection rules
and requirements for non-vanishing asymmetry in the resulting angular distribution when averaging over all orientations. 

The anisotropy parameters $\beta^{1ph}_{L,M}(\epsilon_{k})$, associated with the one-photon ionization pathway, cf.~Appendix~\ref{sec:appendix2}, 
can be expressed as,
\begin{subequations}
\begin{eqnarray}
\label{eq:beta1.final} 
  \beta^{1ph}_{L,M}(\epsilon_k)&=& \sum_{\mu_0,\mu^\prime} c^{(1ph)}_{\mu_0,\mu^\prime}(L)
\begin{pmatrix} 1 & 1 & L\vspace*{0.31cm}\\ -\mu_0 & \mu^\prime_0 &-M\end{pmatrix}\,.  
\end{eqnarray}
  The exact form for the coefficients $c^{(1ph)}_{\mu_0,\mu^\prime}(L)$ are given in 
Appendix~\ref{subsubsec:beta11}.
In particular for linearly polarized fields i.e.,  $\mu_0=\mu^\prime_0=0$, the Wigner $3j$-symbol in Eq.~\eqref{eq:beta1.final} vanishes for $L$ odd. Consequently, 
first order processes do not contribute to the asymmetry. For circularly
polarized light, however, $\mu_0$ and $\mu^\prime_0$ can take values $\pm 1$, which would
lead to non-vanishing contribution for $L=1$ and $M=0$ provided that cancelations upon 
summation
over the bound-continuum dipole matrix elements with opposite sign ``$m$'' magnetic quantum number, here absorbed in the coefficients $c^{(1ph)}_{\mu_0,\mu^\prime}$, does not occur, i.e. for chiral molecules~\cite{RitchiePRA76}. 
The anisotropy parameters associated with second-order processes read, 
\begin{widetext}
\begin{eqnarray}
  \label{eq:beta2.final} 
  \beta^{2ph}_{L,M}(\epsilon_k)&=&
    \sum_{_{\substack{\mu_0,\nu_0\\ \mu^\prime_0,\nu^\prime_0\\Q_1, Q_2}}}c^{(2ph)}_{\scaleto{\mu_0,\nu_0,\mu^\prime_0,\nu^\prime_0,Q_1,Q_2}{7pt}}(L) \begin{pmatrix} 1 & 1 & Q_1\vspace*{0.31cm}\\ -\mu_0 & -\nu_0 & \mu_0+\nu_0 \end{pmatrix}\!\!\!
    \begin{pmatrix} 1 & 1 & Q_2\vspace*{0.31cm}\\ -\mu^\prime_0 & -\nu^\prime_0 & \mu^\prime_0+\nu^\prime_0 \end{pmatrix}\!\!\!
    \begin{pmatrix} Q_1 & Q_2 & L\vspace*{0.31cm}\\ -\mu_0-\nu_0 & \mu^\prime_0+\nu^\prime_0& -M\end{pmatrix}\,.\quad\quad 
 \end{eqnarray}
\end{widetext}
Derivation and explicit form for the coefficients $\beta^{2ph}_{L,M}(\epsilon_k)$ are detailed in 
Appendix~\ref{subsubsec:beta22}. The selection rules for two-photon process are analogous to that described for
the one-photon counterpart. In particular, 
the third Wigner symbol in Eq.~\eqref{eq:beta2.final} vanishes for $L$ odd
for linearly polarized fields as the first and second Wigner symbols vanish
for odd $Q_1$ and $Q_2$.

  Finally, in Appendix~\ref{subsubsec:beta12}, we show that the laboratory-frame orientation-averaged anisotropy parameter associated with the interference between both
photoionization pathways, defined in Eq.~\eqref{eq:beta.int.def},  has the following structure,
 \begin{widetext}
 \begin{eqnarray}
 \label{eq:beta12.final} 
   \beta^{int}_{L,M}(\epsilon_k)&=&
    \sum_{Q_1,Q_2}\sum_{\substack{\mu_0,\mu^\prime_0}}\sum_{\nu^\prime_0}c^{(int)}_{\scaleto{\mu_0,\mu^\prime_0, \nu^\prime_0,Q_1,Q_2}{7pt}}(L)\begin{pmatrix} 1 & 1 & Q_2\vspace*{0.31cm}\\ -\mu^\prime_0 & -\nu^\prime_0 & \mu^\prime_0+\nu^\prime_0 \end{pmatrix}\!\!\!
   \begin{pmatrix} 1 & Q_2 & L \vspace*{0.31cm}\\ -\mu_0 & \mu^\prime_0+\nu^\prime_0 & -M \end{pmatrix}\!.\nonumber\\ 
 \end{eqnarray}
 \end{widetext}
\end{subequations}
In contrast to Eqs.~\eqref{eq:beta1.final} and~\eqref{eq:beta2.final}, the interference between one- and two-photon ionization pathways
may lead to non-vanishing anisotropy parameters for $L$ odd when the driving field is linearly polarized. This feature
persists even after the orientation averaging. In fact,
the second Wigner symbol in Eq.~\eqref{eq:beta12.final} does not vanish for
$L$ odd and $Q_2$ even, for $M=0$, when $\mu_0=\mu^\prime_0=\nu^\prime_0=0$. Even values for $Q_2$ are allowed
by the first Wigner $3j-$ symbol. In the following, we describe our
optimization approach to manipulate the anisotropy parameters using different
photoionization schemes and polarization configurations in the quest to maximize 
the anisotropy in the photoelectron emission.

\subsection{Control of the photoionization dynamics}

In order to control the photoionization dynamics, we  consider coherent control of wave packet interference mediated by linearly polarized 
or polarization shaped pulses. In the first instance, the pulse is assumed to
 be linearly polarized --parallel to the $\text{\textbf{e}}^\prime_z=\text{\textbf{e}}^\prime_0$ axis-- 
 and defined as a coherent superposition of $N$ sub-pulses,  
  \begin{subequations}
  \label{eq:parametrization}
  \begin{eqnarray}
  \label{eq:parametrization1}
    \text{\textbf{E}}(t)\cdot\text{\textbf{e}}^\prime_0 = \mathscr{E}^\prime_{0}(t)&=&\sum^{N}_{j=1} \mathscr{E}^\prime_{0,j}(t)\,, 
\end{eqnarray}
    where $\mathscr{E}^\prime_{0,j}(t)$ is the sub-pulse carrying the frequency $\omega_j$
and parametrized according to
  \begin{eqnarray}
  \label{eq:parametrization2}
    \mathscr{E}^\prime_{j,0}(t)&=&h_j(t-\tau_j)\,\cos\Omega_j(t)\,, 
\end{eqnarray}
with $\Omega_j(t) =\omega_j(t-\tau_j) + \phi_j $  and where $h_{j}(\cdot)$ is a Gaussian envelope of the form      
\begin{eqnarray}
  h_j(t-\tau_j) = \text{E}_{0,j}\times e^{-(t-\tau_j)^2/2\sigma_j}\,.
\end{eqnarray}
\end{subequations}
The pulse parameters $\text{E}_{0,j}$ $\omega_{j}$, and $\phi_{j}$  
define the peak field amplitude,
central frequency, and carrier envelope phase of the $j$th sub-pulse with full width at half 
maximum $\text{FWHM}=2\sqrt{2\ln{2}}\sigma_{j}$ whose peak intensity is delayed by $\tau_{j}$ with respect to $t=0$.

For polarization-shaped fields, we consider pulses with circular right (CRP) and left (CLP) rotating polarization 
directions and define the driving field  as a linear combination
thereof, 
  \label{eq:combinationRL}
  \begin{eqnarray}
    \text{\textbf{E}}^{\prime}(t) =\text{\textbf{E}}^{\prime}_{\text{R}}(t)+\text{\textbf{E}}^{\prime}_{\text{L}}(t). 
\end{eqnarray}
The CRP and CLP components are defined from the point of view of the emitter and
parametrized following the guidelines detailed  
in Appendix~\ref{sec:polarization_param}.

For further inspection of the electron dynamics driven by  polarization shaped pulses, we  define  $\zeta_j(t)$ as 
the helicity of the sub-pulse %defined in Eq.~\eqref{eq:parametrization1} 
carrying  the frequency $\omega_j$, which we write in terms of the differential quantity~\cite{guenther1990interference,hockett2015coherent}    
\begin{eqnarray}
  \label{eq:zeta.def}
  \zeta_j(t)&=&\dfrac{|\text{\textbf{E}}^\prime_{R,j}(t)|-|\text{\textbf{E}}^\prime_{L,j}(t)|}{|\text{\textbf{E}}^\prime_{R,j}(t)|+|\text{\textbf{E}}^\prime_{L,j}(t)|}\,,
\end{eqnarray}
with $\text{\textbf{E}}^\prime_{R,j}(t)$ $(\text{\textbf{E}}^\prime_{L,j}(t))$ the portion of the field with CRP (CLP) 
carrying the frequency component $\omega_j$.

The photoelectron observable $\mathcal{I}(\epsilon,\theta_{\boldsymbol{k}^\prime})$ 
is an energy- and angle-resolved measurable quantity proportional to photoelectron probability distribution
defined in Eq.~\eqref{eq:SecondOrderApprox_split} and given by~\cite{goetz2019quantum},
\begin{eqnarray}
  \label{eq:Idef}
  \mathcal{I}(\epsilon,\theta_{k^\prime})&\propto&
  \left.\dfrac{d^2\sigma}{d\epsilon_k\,d\Omega_{\boldsymbol{k}^\prime}}\right\vert_{\phi_{\boldsymbol{k}^\prime}=\pi/2}. 
\end{eqnarray}
We define the 
intensity-normalized anisotropy of the PAD as
\begin{eqnarray}
  \label{eq:MyA}
  \mathcal{A}(\epsilon_k,\theta_{k^\prime}) = 
  \dfrac{\mathcal{I}(\epsilon_k,\theta_{k^\prime}) - \mathcal{I}(\epsilon_k,\pi-\theta_{k^\prime})}{\mathcal{I}_0}\,,
\end{eqnarray}
where $\theta_{\boldsymbol{k}^\prime}$ is defined by the photoelectron direction
of emission with respect to the light propagation direction for circularly
polarized light or with respect to the light polarization direction for linearly
polarized fields and where $\mathcal{I}_0$ corresponds to the photoelectron peak intensity 
\begin{eqnarray}
 \label{eq:I0}
 \mathcal{I}_0 = \underset{\epsilon_k, \theta_{\boldsymbol{k}^\prime}}{\mathrm{max}}\,\mathcal{I}(\epsilon_k,\theta_{\boldsymbol{k}^\prime})\,. 
\end{eqnarray}
Next, we define the optimization problem by
\begin{eqnarray}
\label{eq:octproblem}
  \operatorname*{arg\,max}_{\text{\textbf{E}}^{\prime}(t)\in U} \left\{\underset{\epsilon_k, \theta_{\boldsymbol{k}^\prime}}{\mathrm{max}}\,\Big| \mathcal{A}(\epsilon_k,\theta_{\boldsymbol{k}^\prime})\Big|\right\}\,,
\end{eqnarray}
where $U$ is the subset of feasible solutions, i.e. constraints that the
parameters defining each component of the electric field $\text{\textbf{E}}^\prime(t)$
must fulfill such as maximal peak intensity, maximal duration (FWHM),  allowed frequency components or
maximal time-delay between two distinct frequency components. The functional form 
of the driving field is parametrized according to
Eqs.~\eqref{eq:parametrization} for linearly polarized fields and using Eqs.~\eqref{eq:pulseparamR} and~\eqref{eq:pulseparamL}
for fields with time dependent helicity. These parameters are optimized using a  
gradient-free sequential update-based method detailed in Ref.~\cite{GoetzSpa2016}
to maximize the anistropy of photoelectron emission probability as defined in
Eq.~\eqref{eq:octproblem}.

Throughout the text, the term anisotropy will be used to refer to %defined 
the quantity defined in
Eq.~\eqref{eq:MyA},  which will be expressed in percentage (of $\mathcal{I}_0$). Perfect anisotropy is
thus only obtained when, for some optimal kinetic energy $\epsilon^*_k$ and emission angle
$\theta^{*}_{\boldsymbol{k}^\prime}$, the anisotropic
component of the photoelectron signal corresponds to the peak intensity
$\mathcal{I}_0$. 

\section{Numerical Results}
\label{sec:NumericalResults}

\subsection{Photoelectron anisotropy with linearly polarized fields}
\label{subsec:CoherentControlBichro}

\begin{figure}[!tb]
  \centering
  \includegraphics[width=0.70\linewidth]{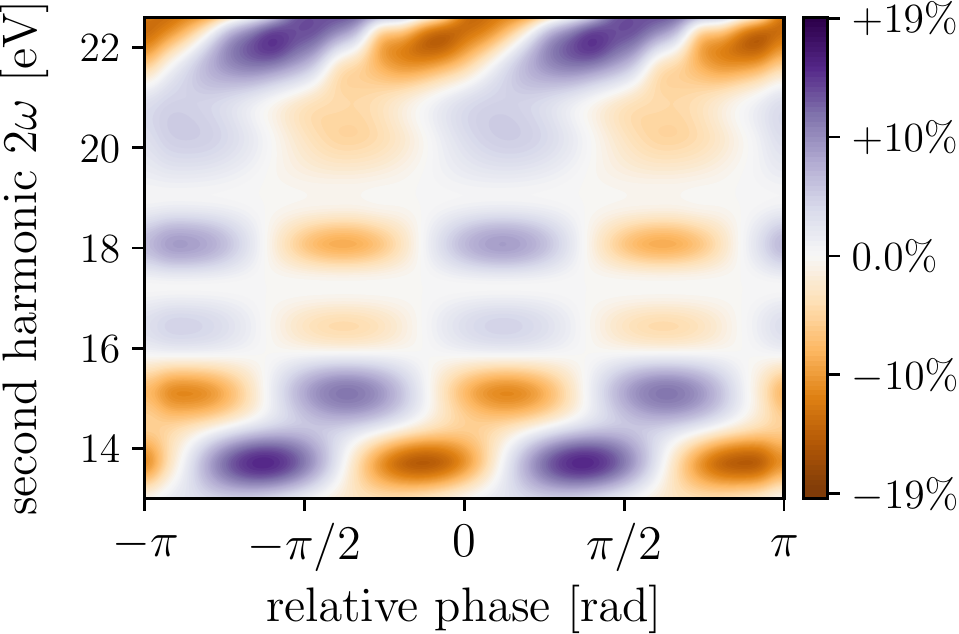}
  \caption{                          
   Left-right asymmetry in the PAD as a function of the photon energy and phase difference between
  the fundamental and second harmonic of a
  linearly polarized bichromatic $(\omega, 2\omega)$ pulse.  
  }
 \label{fig:Figure01}
\end{figure}

We start by considering non-sequential phase-controlled bichromatic $(\omega, 2\omega)$
ionization from a randomly oriented ensemble of $\ce{CHBrClF}$ molecules
driven by linearly polarized fields, with polarization direction parallel to the $z^\prime-$axis. 
Figure~\eqref{fig:Figure01} shows 
the left-right anisotropy of photoelectron emission as a function of the photon
energy (second harmonic) and relative phase between both colors. This corresponds to the typical scenario discussed in
Refs.~\cite{Muller1990,YinPRL92,SchumacherPRL94,gryzlova2018quantum,douguet2016photoelectronCirc,grum2015photoelectron,GrumPRA2015,gryzlova2018quantum}.
The asymmetries shown in Fig.~\eqref{fig:Figure01} were obtained by considering a $1:2$ ratio of 
fundamental to second harmonic,  with $I_0=10^{11}\,$W/cm$^2$ for the fundamental.
A pulse with a width of $23\,$fs was used for both colors. 
This ratio leads to comparable ionization yields 
from both pathways, inducing a noticeable break of symmetry in the PAD, which
is independent of the chiral nature of the target~\cite{shapiro2012quantum,douguet2016photoelectronCirc,GrumPRA2015,grum2015photoelectron} as the field
is linearly polarized. 
The anisotropy originates from a coherent wave packet interference 
between    
single- and two-photon  photoionization
pathways. 
Periodic oscillations of the
anisotropy as a function of the relative phase can be observed, confirming the
coherent nature of the control mechanism. Overall, the
left-right anisotropy exhibits moderate values not exceeding $\pm 20\%$ 
for the chosen field parameters.  
In what follows, we will discuss the efficiency and limitations of bichromatic coherent control 
for achieving perfect anisotropy in randomly oriented photoionized molecules.

\begin{figure}[!tb]
  \centering
  \includegraphics[width=0.99\linewidth]{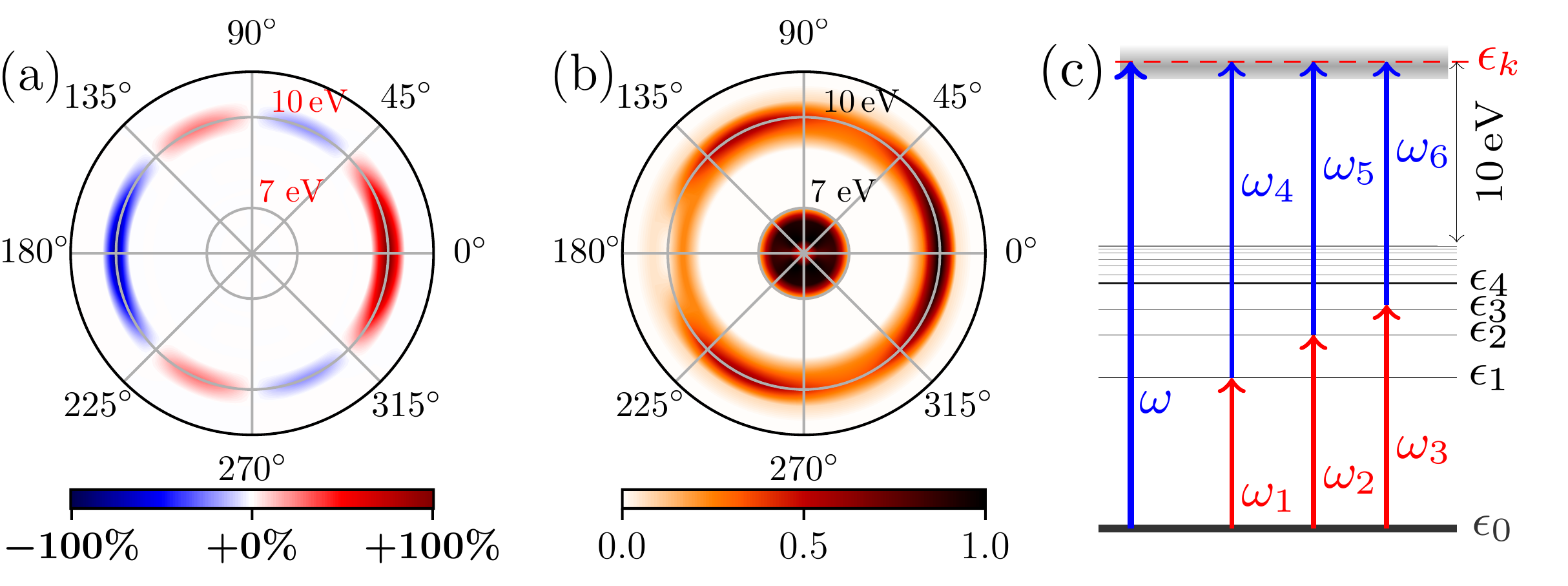}
  \caption{                          
   (a) Optimized anisotropy in the PAD corresponding to the 
   multiple-REMPI scheme, achieving $100\%$ of asymmetry at a photoelectron kinetic
   energy of $10\,$eV and angles $\theta_{\boldsymbol{k}^\prime}=180$ and $\theta_{\boldsymbol{k}^\prime}=0$ degrees.  (b)
   Corresponding photoelectron momentum distribution with zero probability of
   photoelectron emission at $\theta_{\boldsymbol{k}^\prime}=180^o$ and maximal ionization probablity
   at $\theta_{\boldsymbol{k}^\prime}=0^o$. (c) Schematic energy representation
   of the photoionization process leading to the observed asymmetry.
  }
 \label{fig:Figure03}
\end{figure}

\begin{figure}[!tb]
  \centering \includegraphics[width=0.95\linewidth]{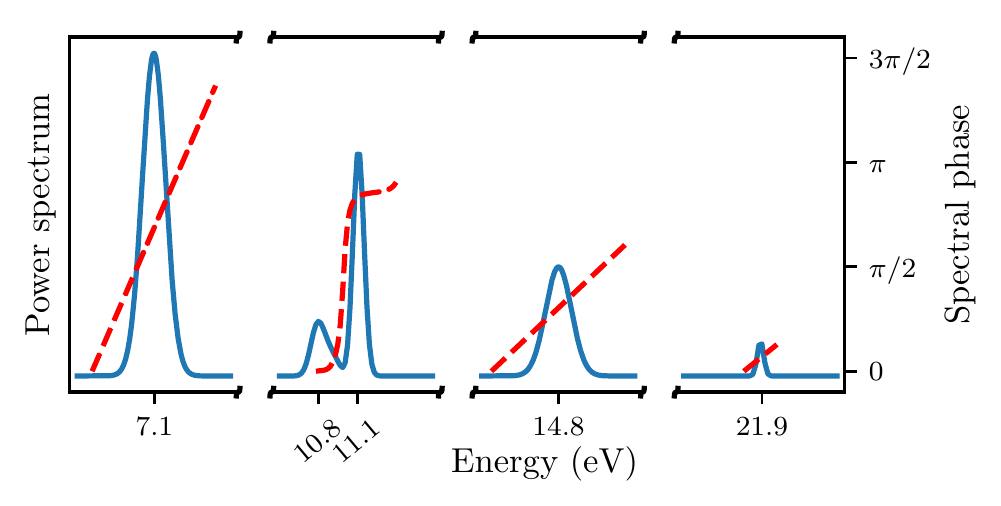}
  \caption{Optimized electric field spectrum (blue) and spectral phase (dashed
    red line) leading to the left-right
  anisotropy of photoelectron emission of $100\%$ shown in Fig.~\ref{fig:Figure03}. 
  The optimal field contains the required photon energies to  generate
  two-photon (7.1-14.8 eV) and single-photon (21.9 eV) pathways 
  that constructively interfere at $10\,$eV, as schematized in 
  Fig.~\ref{fig:Figure03}(c): 
  The low frequency component at $7.1\,$eV in 
  resonantly excites the LUMO. The peaks at 10.8 and 11.1 eV --within their
  bandwidth-- resonantly excite the LUMO+$j$ orbitals, for $j>1$. The energy
  required for ionization of the resonantly exited LUMO orbitals at
  a photoelectron kinetic energy of $10\,$eV  is available
  within the spectral bandwidth around the peak at $14.8\,$eV. The 21.9 eV
    frequency is responsible for
  single-photon ionization of the LUMO at a photoelectron kinetic energy of
  $10\,$eV.} 
 \label{fig:Figure05}
\end{figure}

To answer the question whether perfect anisotropy ($100\%$) 
in a randomly oriented sample of molecules is achievable by
coherent control using suitably shaped  
ionizing pulses beyond the non-sequential bichromatic case,
we optimize multi-color fields, defined in Eq.~\eqref{eq:parametrization}, 
first constraining the polarization state to linear  
and the maximal peak intensity to not exceed $1.0\times 10^{12}\,$W/cm$^2$.
This intensity threshold has been found to be an appropriate upper limit for
the validity of the perturbation treatment in bichromatic 
photoionization studies~\cite{douguet2016photoelectronCirc}.  
Figure~\ref{fig:Figure03}(a) displays the left-right asymmetry in the PAD obtained upon
optimization of the linearly polarized multi-color field. The corresponding PAD,
shown in Fig.~\ref{fig:Figure03}(b), exhibits perfect
left-right anisotropy $(100\%)$ exactly at a photoelectron kinetic energy of $10\,$eV 
with maximal probability of photoelectron emission parallel
to the field polarization axis and minimal probability of emission
anti-parallel to this axis.
In order to investigate the coherent mechanism leading to the  
anisotropy of $100\%$ displayed in Fig.~\ref{fig:Figure03}(a), we analyze the optimized frequency components and spectral phases 
in Fig.~\ref{fig:Figure05}. 
In detail, the photon energy distributions shown in Fig.~\ref{fig:Figure05} at
$7.1\,$eV and $14.8\,$eV
ensure resonant photoionization of the HOMO --through the LUMO--  to a final photoelectron kinetic energy of $10\,$eV. 
The lowest photon energy of $7.1\,$eV 
corresponds to the resonant transition energy between the HOMO and LUMO, with
orbital  energies corresponding, at the aug-cc-pVDZ level, 
to $-11.878\,$eV and $-4.803\,$eV, respectively.   
Interestingly, the double-peaked photon energy distribution shown in
Fig.~\ref{fig:Figure05} at $10.8\,$eV and $11.1\,$eV has
a four-fold purpose with nested contributions to the excitation-ionization
steps: It contains
\begin{enumerate}[label=(\alph*)]
  \item\label{enum:a}  
the required photon energy of $10.97\,$eV (first peak) to excite the  
    transition $\mathrm{HOMO}\rightarrow\mathrm{LUMO}+1$ 
\item\label{enum:b}  
the photon energy of $11.063\,$eV (second peak) to resonantly excite 
     the transition  $\mathrm{HOMO}\rightarrow\mathrm{LUMO}+2$,\label{item:b} 
 \item\label{enum:c}  
the appropriate photon energy of $10.97\,$eV (second peak) to ionize the  LUMO$+1$ to a photoelectron
kinetic energy of exactly $10\,$eV,  
   \item\label{enum:d}   
and, within the spectral distribution around $10.8\,$eV, 
     the  photon energy of $10.814\,$eV to ionize the  LUMO$+2$ (first peak)
     also at $10\,$eV. 
\end{enumerate} 
Finally, the spectrum of the optimized field also
contains the photon energy of $21.875\,$eV--at an intensity of $4.11\times 10^9$W/cm$^2$--  
which is required
to ionize the HOMO at a photoelectron kinetic energy of
$10\,$eV. Simultaneous removal of the frequency
components around $(i)$ $7.1\,$eV and $10.8-11.1\,$eV in Fig.~\ref{fig:Figure05} 
or $(ii)$ those around  $10.8-11.1\,$eV and $14.8\,$eV or $(iii)$ that centered at $21.9\,$eV alone
results in zero left-right asymmetry. In case $(i)$ and $(ii)$, 
the required photon energies to induce resonantly-enhanced two-photon ionization 
at a photoelectron energy of $10\,$eV probing the first three lowest unoccupied molecular orbitals 
are inaccessible. In $(iii)$, these even-parity
photoionization pathways are enabled but the odd-parity pathway is disabled.
For linearly polarized light, interferences between same-parity photoionization
pathways do not break the asymmetry as discussed in Sec.~\ref{subsec:anisotropyparameters}. Consequently, no anisotropy is observed in $(iii)$.
Conversely, removing only the photon energies of $10.8-11.1\,$eV, which induce resonant
ionization probing the LUMO+1 and LUMO+2, or those corresponding to 
$7.1\,$eV and $14.8\,$eV, which promote resonant ionization through the LUMO, results in non-vanishing anisotropy at $10\,$eV. 
These observations suggest an control mechanism based on
coherent wave packet interferences mediated by one-photon ionization and a manifold
of two-photon ionization pathways.  Furthermore,  altering the
spectral phase shown in Fig.~\ref{fig:Figure05} while keeping the power spectrum  unchanged, dramatically alters the resulting asymmetry, leading to 
significantly smaller magnitudes (below $10\%$, depending on the spectral phase modifications), confirming
the coherent nature of the control mechanism. We therefore conclude that the enhancement mechanism    
is mediated by constructive quantum interferences between the different portions of the
coherent photoelectron wave packet resulting from the odd-parity single-photon ionization
channel and a manifold of even-parity resonant ionization pathways involving the 
first three molecular excited states. 

Asymmetries in the PADs  
are well-known to be
sensitive to the photoelectron kinetic energy, see e.g., refs.~\cite{gryzlova2018quantum,GrumPRA2015,gryzlova2018quantum}
for bichromatic ionization in linearly polarized fields. In order to disentangle the contributions from the final continuum state (here
defined by the continuum state with energy $10\,$eV at which the multiple-REMPI achieves
perfect anisotropy) and
those originating from the ionization pathways (defined by the multiple-REMPI paths), we  
optimize linearly polarized fields to maximize the left-right asymmetry at the same photoelectron kinetic energy of
$10\,$eV, but
constraining the optimized pulse spectrum to bichromatic $(\omega,2\omega)$ components.
This corresponds to (fixed) photon energies carried by the fundamental and second harmonic  
of $\omega=10.939\,$eV and  $2\omega=21.878\,$eV, respectively. It is worth
noticing that both control approaches, namely multiple-REMPI and bichromatic schemes,  share the photon
energy of $2\omega=21.878\,$eV and both being optimized at the same photoelectron
kinetic energy, any difference in the outcome is thus solely due to
an intermediate-pathway effect. 

\begin{figure}[!tb]
  \centering
  \includegraphics[width=0.90\linewidth]{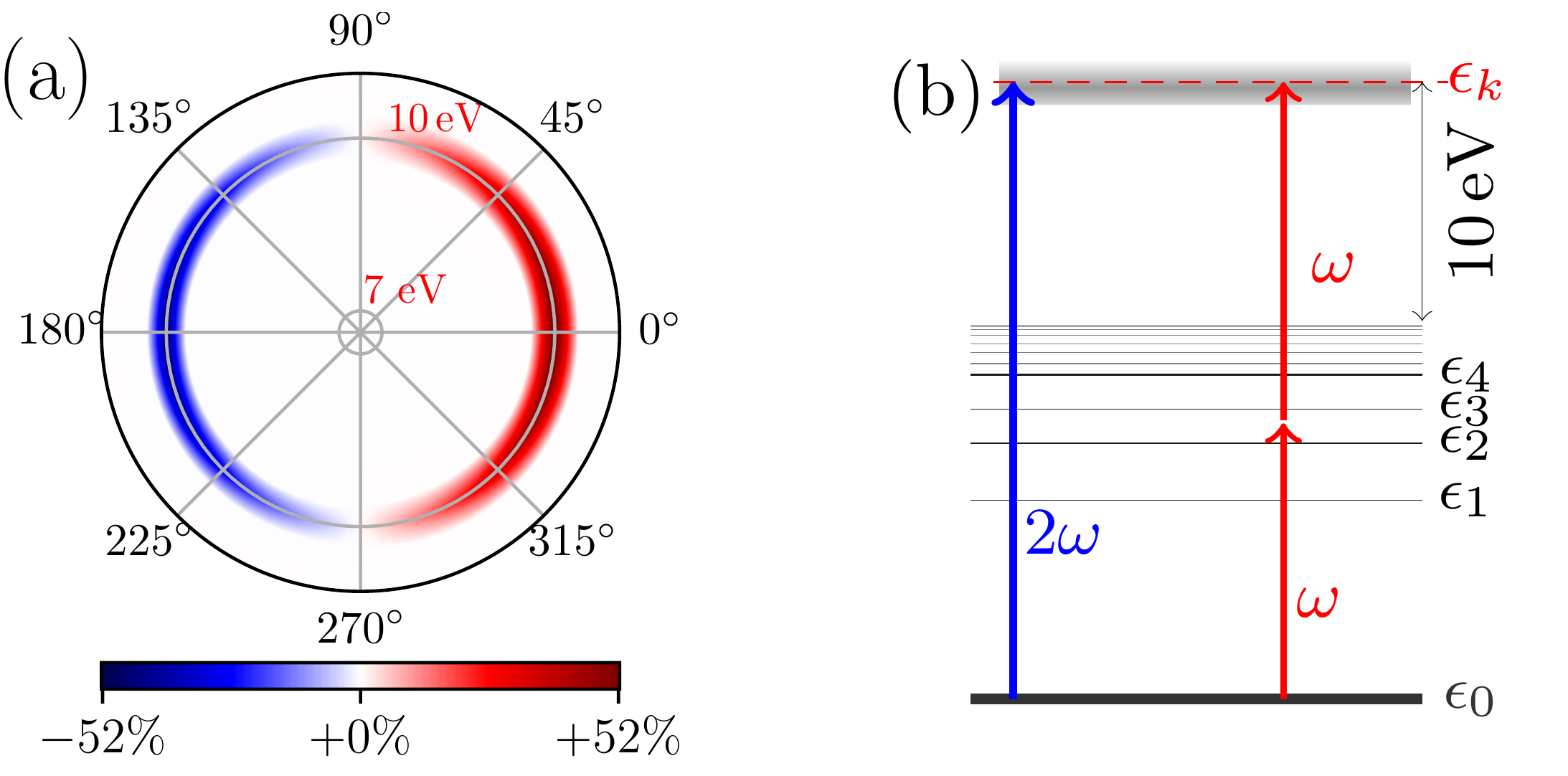}
  \caption{                          
  (a) For a photoelectron kinetic energy of $10\,$eV, a maximal anisotropy of $52\%$ is obtained with the optimized
  bichromatic $(\omega,2\omega)$, in contrast to $100\%$ for the Multiple-REMPI
  scheme shown in Fig.~\ref{fig:Figure03}. (b) Photoionization scheme for the
  optimized bichromatic $(\omega, 2\omega)$ pulse.
  }
 \label{fig:Figure06}
\end{figure}

Figure~\ref{fig:Figure06}(a) shows the maximal achievable anisotropy at
a photoelectron energy of $10\,$eV obtained with the linearly
polarized optimized bichromatic $(\omega, 2\omega)$ pulse.  With 
a maximal left-right asymmetry  of $52\%$ at $10\,$eV
the performance of the bichromatic $(\omega, 2\omega)$ ionization scheme is significantly inferior to the
multiple-REMPI scheme. The smaller asymmetry obtained in the
bichromatic scenario can be explained by the fact that resonant excitation driven by the fundamental is not fully achieved. In fact, 
the two-photon pathway is in resonance at $-0.939\,$eV, 
which lies between the orbital energies of  LUMO$+1$ ($-0.974\,$eV) and LUMO$+2$ ($-0.8136\,$eV) orbitals, as
depicted in~Fig.~\ref{fig:Figure06}(b). 

\begin{figure}[!tb]
  \centering
  \includegraphics[width=0.75\linewidth]{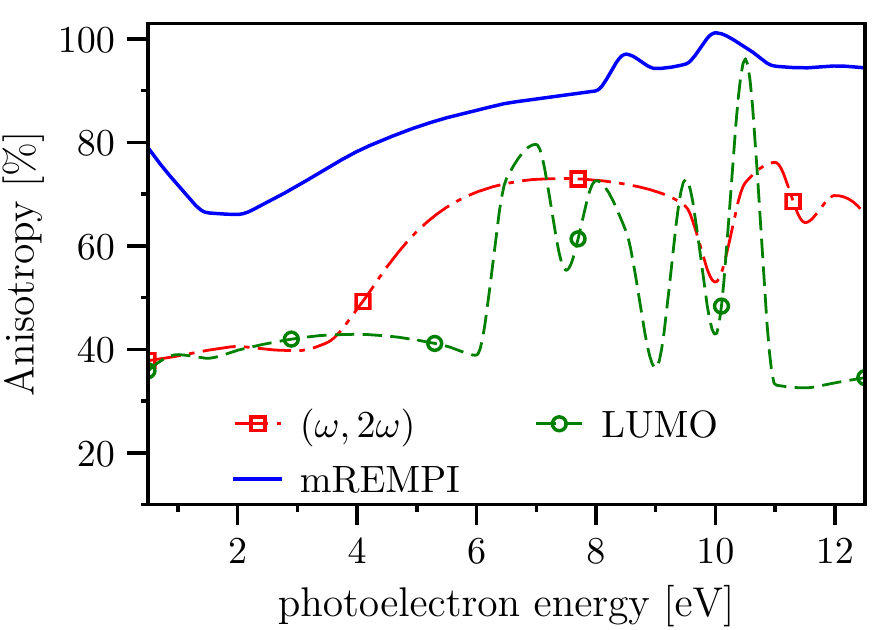}
  \caption{                          
  Maximal achievable anisotropy for 
  different photoionization schemes: linearly polarized bichromatic $(\omega, 2\omega)$ pulse (dot-dashed red line),
    multiple-REMPI (filled blue line) and 3-color (LUMO) case (dashed-green) discussed in the text.  
  }
 \label{fig:Figure02}
\end{figure}

Therefore, for an objective comparison between the bichromatic and multiple-REMPI approaches,
we further optimize linearly polarized fields using both schemes but within a range of different
photoelectron kinetic energies. Figure~\ref{fig:Figure02} displays the maximal
achievable left-right asymmetry obtained with both, the multiple-REMPI (solid-blue line)  
and the bichromatic $(\omega, 2\omega)$ (dot-dashed red line) schemes. 
The oscillations in  Fig.~\ref{fig:Figure02} illustrate the sensitivity of the anisotropy to the final  continuum state for the different control schemes.
Nevertheless, and with no 
exception, the multiple-REMPI scheme systematically outperforms the
bichromatic $(\omega, 2\omega)$ counterpart.
%% chr: no new paragraph
  It is worth noticing that decreasing the number of
  (resonant) interfering paths results in an overall decrease in the left-right asymmetry. For instance,  
  the 3-color (LUMO) case shown in Fig.~\ref{fig:Figure02} (dashed-green line) corresponds to
  a particular case of the multiple-REMPI scheme where only a single even-parity
  two-photon pathway, in resonance with the LUMO, interferes with the
  odd-parity ionization channel.

\subsection{Optimal control in multi-color polarization-shaped fields}
\label{subsec:polarization.shaped}

Extension of the multiple-REMPI scheme to circularly-polarized fields
provides an additional degree of freedom for the possible interfering pathways. 
Here, quantum interferences between opposite-parity and even-parity two-photon ionization paths can 
be exploited to exert control over the forward-backward asymmetry in the
photoelectron emission probability~\cite{goetz2019quantum}. We
find that fixing the field polarization state to either left- or right-circular for all optical pathways leads to a maximal forward-backward
asymmetry of $68\%$ at $6.5\,$eV and $64\%$ at $10\,$eV. These results clearly indicate that the
orientation-averaged asymmetry in PAD is sensitive to the details of the polarization of the driven field. 

\begin{figure}[!tb]
  \centering
  \includegraphics[width=0.80\linewidth]{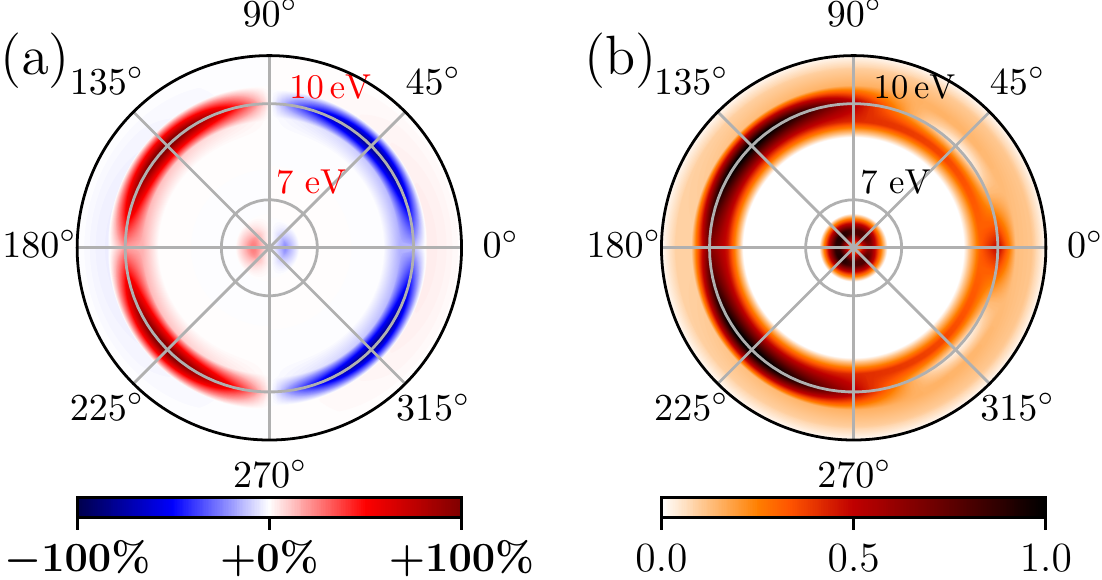}
  \caption{                          
  (a) Asymmetric component in the forward-backward photoelectron emission
  probability obtained with the optimized multi-color
  field with time-dependent helicity. The field  have been optimized to
  maximize the anisotropy of photoelectron emission at a photoelectron kinetic energy
  of $10$\,eV. Corresponding photoelectron momentum distribution shown in (b). 
  }
 \label{fig:FigureM1}
\end{figure}

\begin{figure}[!tb]
  \centering
   \includegraphics[width=0.90\linewidth]{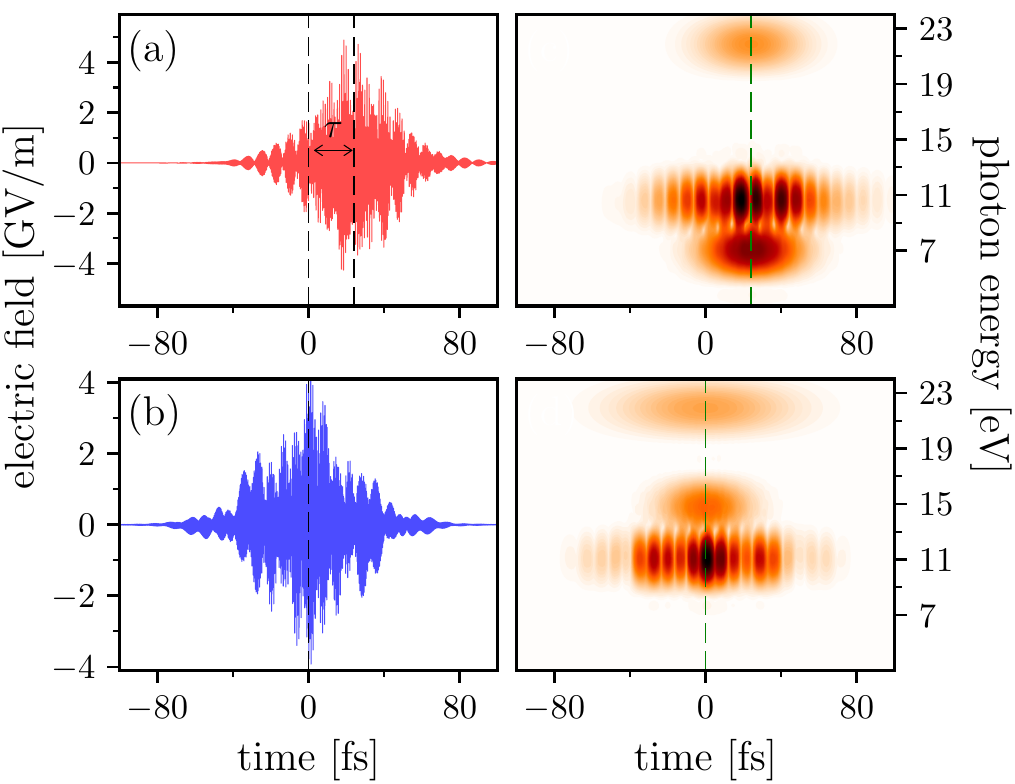}
  \caption{                          
  Projection of the circularly right (a,c) and left (b,d) rotating components of the optimized field into the $x^\prime-$ axis.
  A time delay $\tau\approx 24\,$fs between both counter-rotating components is observed in
  the temporal, panels (a) and (b), dashed black lines, as well as in the frequency domain, cf. panels (c) and (d), dashed white-lines. However,
  within a given rotating direction, left or right, all frequency components are
  perfectly synchronized.
  }
 \label{fig:FigureM2new}
\end{figure}

It is nevertheless possible to retrieve 
perfect forward/backward anisotropy, i.e. $100\%$, in the direction of  photoelectron emission  
by shaping the polarization state of the driving field in time.
In other words, we render the helicity of the field polarization time-dependent.
This can be achieved by introducing 
different pulse durations, phases and time delays 
to the pulses with projection in counter-rotating directions~\cite{hockett2015coherent}. 
\label{subsec:polshaping}
We test this approach to maximize the forward-backward photoelectron emission probability at
a photoelectron kinetic energy of $10\,$eV. This energy corresponds to the
photoelectron kinetic  
energy at which a perfect anisotropy of $100\%$ was obtained using 
the optimized linearly polarized pulse, cf.~Fig.~\ref{fig:Figure03}. 
Figure~\ref{fig:FigureM1}(a) displays the resulting asymmetric component of the
PAD.  The optimized 
forward-backward anisotropy amounts to $100\%$ at the kinetic photoelectron energy $\epsilon^*_k$ of $10\,$eV 
along the direction $\theta^*_{\boldsymbol{k}^\prime}=135^\circ$. The optimized momentum distribution
shown in Fig.~\ref{fig:FigureM1}(b)
exhibits vanishing emission probability in the direction defined by  $\theta_{\boldsymbol{k}^\prime}=45^\circ$ and maximal
photoemission probability at $135^\circ$ for $10\,$eV. Here, 
$\theta_{\boldsymbol{k}^\prime}$ corresponds to the polar angle with respect to the light propagation direction --assumed to define 
the $z^\prime-$axis 
-- and corresponding to $\theta_{\boldsymbol{k}^\prime}=0^\circ$. The optimized field with
time-dependent helicity is shown in Fig.~\ref{fig:FigureM2new}, with panels (a)
and (b) showing the optimized circularly right and left rotating components respectively.
The multicolor field with time-dependent helicity  is able to reach a
forward-backward asymmetry of $100\%$, in contrast to $64\%$ reached by optimizing the field with
fixed (left or right) helicity, i.e. time-independent helicity. 

\begin{figure}[!tb]
  \centering
  \includegraphics[width=0.80\linewidth]{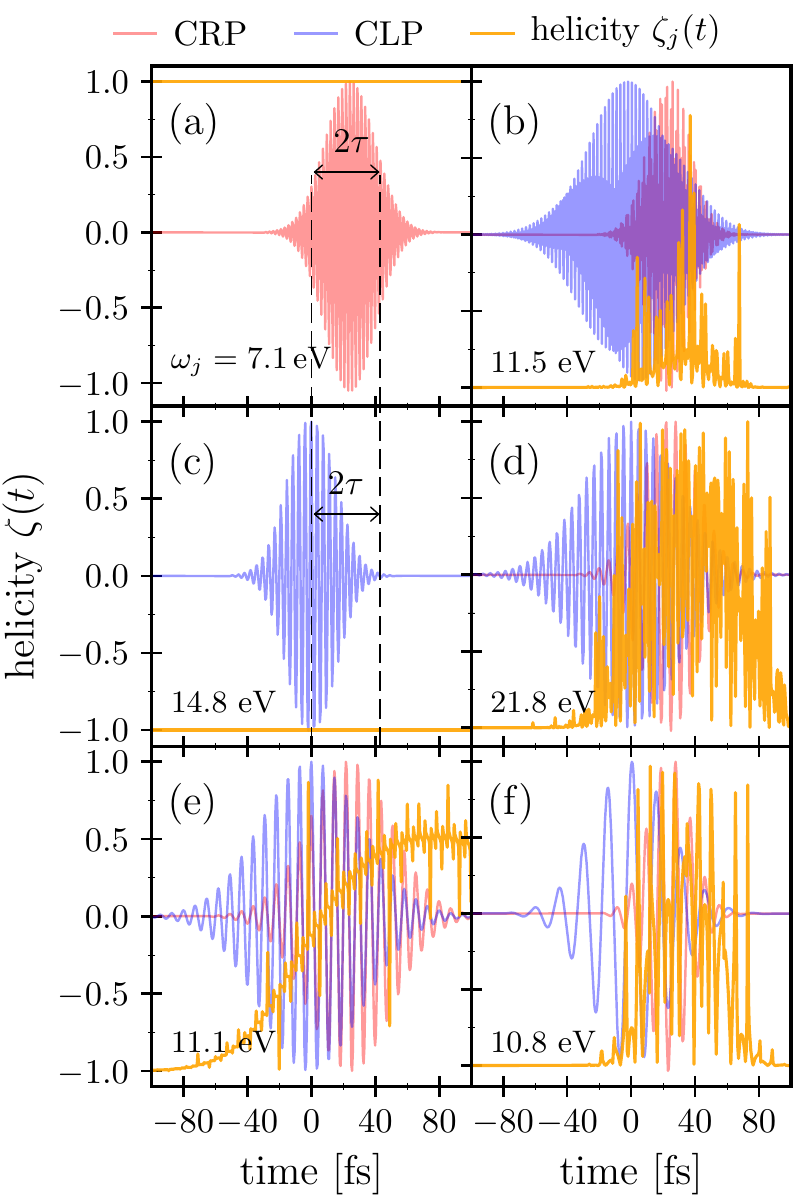}
  \caption{(a)-(f) Electric field helicity $\zeta_j(t)$ for a sub-pulse carrying
  the frequency $\omega_j$ as function of time (solid-orange
  line).   
  The optimized field contains sub-pulses 
  carrying frequencies $\omega_j$ with simultaneous projections along both
  counter-rotating polarization directions.  
  When both counter-rotating components share the same frequency $\omega_j$,  
  the time-delay between the sub-pulses carrying $\omega_j$   has
  a two-fold role: first, it synchronizes the resonant excitation-ionization processes  between
  the contributing resonant photoionization paths
  and secondly, it defines the temporal profile of the helicity (along with
  pulse FWHM, phases, peak intensity)
  to achieve $100\%$ anisotropy by exploiting the sensitivity of the
  PAD to the temporal changes of the field polarization direction, 
   which was not possible when a circularly polarized
  field with constant helicity\, i.e., $(\zeta(t)= \pm 1)\,$, was prescribed to achieve the same goal.
  }
 \label{fig:FigureZETA}
\end{figure}

In order to quantify the main contribution to the enhancement, 
we further inspect the spectral components %and 
for each counter-rotating component. 
Figure~\ref{fig:FigureM2new}(c) and
(d) display the projections onto the $x^\prime-$ axis of the time-frequency distribution
for the circularly right (c) and left (d) rotating fields.
By comparing Figs.~\ref{fig:FigureM2new}(c) and (d), it is apparent that
a non-negligible  
time delay separates the time-frequency distribution along both counter-rotating
directions, which could also already be
noticed in Fig.~\ref{fig:FigureM2new}(a) and (b). Interestingly,  
for a given
rotation direction --left or right-- all frequency components are synchronized with no appreciable
time-delay among them, cf.~Fig.~\ref{fig:FigureM2new}(c) and (d). 
Another remarkable difference between the optimized right and left rotating fields
concerns their spectral distribution. While the optimized rotating fields, $\text{\textbf{E}}^\prime_{R}(t)$,
and $\text{\textbf{E}}^\prime_{L}(t)$ 
share the photon energies of  $10.8\,$eV, $10.9\,$eV, $11.06\,$eV and $21.87,$eV,
cf.~ panels (b) and (d) in Fig.~\ref{fig:FigureM2new},  
the photon energy required for excitation of the LUMO ($7.074\,$eV) is only
contained in the circularly right polarized component, cf.~Fig.~\ref{fig:FigureM2new}(b).
Conversely, the photon energy of $14.803\,$eV --required for the ionization of the
LUMO-- is only present along its 
counter-rotating counterpart, cf.~Fig.~\ref{fig:FigureM2new}(d).  
The Wigner-Ville  distribution function in Fig.~\ref{fig:FigureM2new}(c) and (d) indicates that   
the photon energies required for excitation and ionization of the LUMO 
--$7.05\,$eV and $14.80\,$eV-- share a common time window of about $\approx 60\,$fs --which can also be seen in panels (a) and (c) in Fig.~\ref{fig:FigureZETA}--
suggesting  non-sequential resonant excitation-ionization probing the LUMO  as
  part of the optimal ionization mechanism: the resonant excitation of the LUMO is mediated by the clockwise rotating
component, whereas ionization at a photoelectron kinetic energy of $10\,$eV is
ensured by the counter-clockwise component of the field. 

The electron dynamics involving the LUMO$+1$ and LUMO$+2$ turns out to be more
complicated as both counter-rotating fields share the photon energies centered 
around $10.8\,$eV and $11.06\,$eV, which we recall, have the four-fold
purpose of exciting and ionizing the LUMO$+1$ and LUMO$+2$ at a final
photoelectron energy of $10\,$eV,  as already discussed in the case of linearly
polarized fields in Sec.~\ref{subsec:CoherentControlBichro}.  Indeed, according to the time-frequency distribution in
Fig.~\ref{fig:FigureM2new}(d), the left rotating component induces non-sequential
excitation-ionization of the LUMO$+1$ and LUMO$+2$. Also the circularly right
polarized  component of the field does so, although slightly later. However, since both
counter-rotating components share a  
common time-window,  a rich but complex 
resonant excitation-ionization --probing the LUMO+1 and LUMO+2-- driven by
the portion of the field with time-dependent helicity  occurs. 

Finally, the part of the field carrying a photon energy of $21.878\,$eV -- which induces single-photon ionization of the HOMO 
to a final phoelectron kinetic energy of $10\,$eV -- is also decomposed as a linear combination  
along both counter-rotating directions, resulting in a time dependence for the pulse helicity. In Fig.~\ref{fig:FigureZETA}, 
we show the field helicity for some relevant frequency components of the
optimized pulse: constant helicity $\zeta_j(t)=\pm 1$ for frequency components not shared by both
counter-rotating directions, cf.~panels (a) and (c),  and a highly oscillatory time-profile for the
frequency components simultaneously shared by both polarization
directions, panels (b) and
(d) to (f) in Fig.~\ref{fig:FigureZETA}. In the latter scenario, the electric
field polarization direction evolves in a non-trivial fashion.

The origin of the time-delay between the right and left circularly polarized fields shown in Fig.~\ref{fig:FigureM2new} 
is investigated in 
Fig.~\ref{fig:FigureZETA}(a) and (c), showing the temporal profile 
of the electric field amplitude for the sub-pulses with photon energies of $7.1\,$eV and $14.8\,$eV.  
In fact, it can be observed in Fig.~\ref{fig:FigureZETA}(a) 
that the time at which the pulse carrying the photon energy of
$7.1\,$eV reaches its half maximum (at $t=0\,$fs) is precisely aligned
with the peak position of the ionizing field, cf.~Fig.~\ref{fig:FigureZETA}(c),
which reaches its peak maximum also at $t=0\,$fs. It can also be noticed that the FWHM of the pulse in ~Fig.~\ref{fig:FigureZETA}(a) coincides with
half the overall duration of the ionizing pulse, defined by the time interval between the peak position of the ionizing pulse at $t=0\,$fs 
and the time when the ionizing pulse is over $(t\approx42\,\mathrm{fs})$ in Fig.~\ref{fig:FigureZETA}(c). As a result and under such particular conditions, the 
efficiency  of the resonantly-enhanced two-photon 
ionization  is greatly enhanced, as the transient population of the LUMO is efficiently ionized.

\begin{figure}[!tb]
  \centering
  \includegraphics[width=0.95\linewidth]{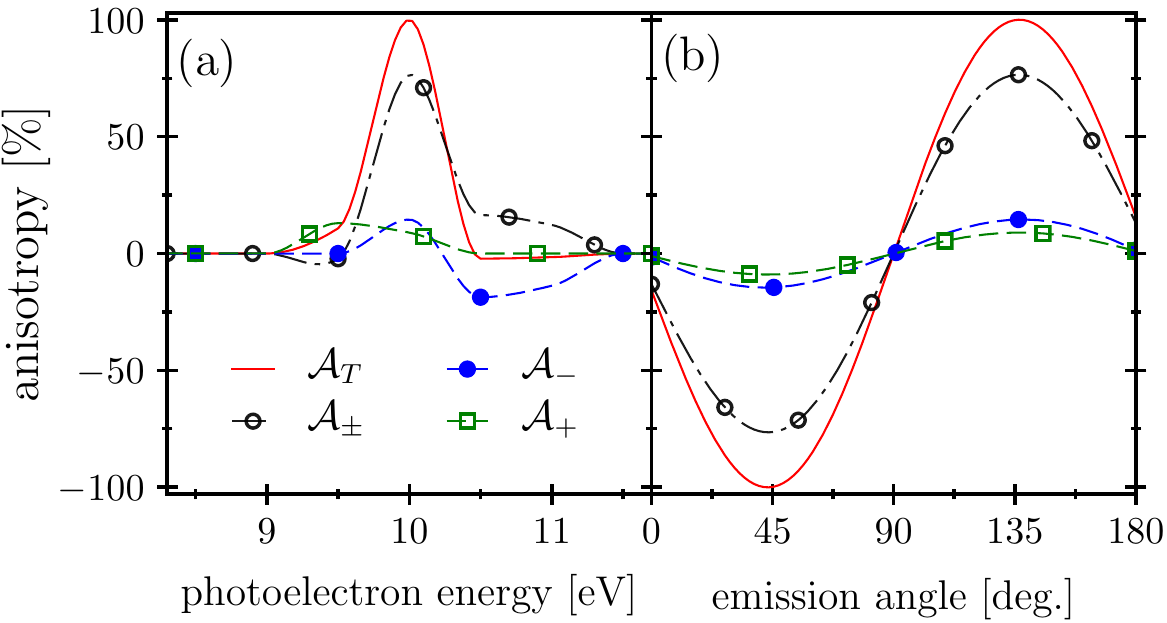}
  \caption{                          
   (a) Anisotropy as a function of the photoelectron energy along the  
   optimal direction of photoelectron emission $\theta^*_{\boldsymbol{k}^\prime}=135^\circ$
   (b) Anisotropy as a function of the emission angle for the optimal photoelectron energy 
   $\epsilon^*_k$ of $10\,$eV.}
 \label{fig:FigureMcartesian}
\end{figure}
Figure~\ref{fig:FigureMcartesian}(a) (solid-red lines) shows the energy-resolved asymmetry along the optimal
direction of photolectron emission $\theta^*_{k^\prime}=135^\circ$ obtained when
both optimized fields components $\text{\textbf{E}}^\prime_{\text{L}}(t)$ and
$\text{\textbf{E}}^\prime_{\text{R}}(t-\tau)$ are simultaneously used for
propagation. This scenario corresponds to the polarization-shaped case, leading to a perfect asymmetry $(100\%)$ at $10\,$eV.  
To further inspect the role of the polarization shaping, we %scrutinize 
examine the partial contribution
to the total anisotropy arising from each counter-rotating component.
This is performed by 
isolating the optimized CRP and CLP 
components of the overall field. Independent propagation then leads to the partial contributions. 
The resulting partial anisotropies are shown in blue filled circles and green empty squares
in Fig.~\ref{fig:FigureMcartesian}(a). 
The individual contribution  from the CRP and CLP rotating   
components accounts for only $15\%$ and $9\%$, respectively.

Remarkably, the partial contributions of each rotating component   
share the same sign at all angles and, in particular, at $135^\circ$, cf.~Fig.~\ref{fig:FigureMcartesian}(b). 
Even more remarkable  is the fact that the leading contribution
arises from coherent wave packet interference that originates from ionization
channels driven  by counter-rotating
components (dotted-dashed black line), which contributes with $76\%$ of the
total asymmetry in the direction $135^\circ$. Such an interference term arises
from the mixed terms involving the product $\propto \mathscr{E}^{\prime}_{+1}\mathscr{E}^{\prime}_{-1}+cc$.
For the optimal set of parameters, the interference term does not vanish upon the orientation averaging. 

The fact that both
counter-rotating components independently contribute with equal sign for the
anisotropy, at the
optimal photoelectron energy, is also
a remarkable feature resulting from the pulse shaping. In fact, 
the asymmetry is expected 
to change sign under ellipticity reversal, e.g. from CRP to CLP,  provided that both rotating components
have the same pulse parameters, i.e. phases, delays, etc.  
However, this is not the case here. 
This suggests a strong interplay between the
optimal phases and time delays of each counter-rotating component that are adjusted by the optimization algorithm  
in such a way that enforces an equal sign for the partial anisotropies
obtained with each individual counter-rotating component, which
further enhances the asymmetry from $76\%$ to $100\%$.

Thus, the isolated contribution from both
counter-rotating directions to the asymmetry amounts modestly, with 24\% of the total
anisotropy, while the contribution from their interference reaches $76\%$. Because such interferences 
are absent in the case of constant $\zeta(t)$ --for which a maximal anisotropy of only $64\%$
at $10\,$eV is obtained-- and owning to the fact that the time-dependence of the
helicity is inherently encoded in the interference term, the high degree of
anisotropy is attributed to the polarization shaping of the ionizing multicolor
field.

\section{Conclusions}
\label{sec:Conclusion}
We have identified two control schemes that achieve perfect anisotropy in a
randomly oriented ensemble of molecules without symmetry. Bichromatic control can
achieve anisotropy in the PAD even after 
orientational averaging, however its efficiency  
to achieve perfect anisotropy was found to be limited. By extending the two-pathway control 
approach to a resonantly-enhanced multiphoton ionization-based control
formalism, we are able to recover full control of the photoelectron dynamics. 
The REMPI scheme
involves interferences between  
odd-parity single-photon ionization pathway and a  
manifold 
of even-parity resonantly enhanced two-photon ionization pathways, 
which  
probe  
different molecular orbitals.
We have shown that for linearly polarized fields,  
the control scheme based on multiple-REMPI 
outperforms bichromatic control for all photoelectron energies.

By generalizing the multiple-REMPI approach to polarization-shaped fields
with time-dependent helicity, we have shown that the forward-backward anisotropy 
in the PAD can also be significantly enhanced. This is achieved by controlling the temporal profile 
of the field helicity. 
The control mechanism is based on  
interference within
a manifold of photoionization pathways driven by fields with counter-rotating polarization directions. Interestingly, %%% est: changed. %%% chr: it is not paradoxical, just a signature of the role of the interference
the isolated contributions of each optimized counter-rotating component produces only  
relatively moderate anisotropy. We have shown that
perfect anisotropy $(100\%)$ is only achieved when both optimized
counter-rotating components are utilized simultaneously.

With the ability to achieve perfect anisotropy in small molecules, we envision
using the multiple-REMPI scheme as a sensitive probe of electron dynamics. Further design
of the pulses would be advantageous, for example to differentiate 
between long-lived dynamic species and multiple product channels. 
In this way,
different measures of the anisotropy of the PADs of complex molecules can be
designed to reveal their complex molecular dynamics. % along predetermined paths in their
We also foresee extending this control procedure to more complex
systems and pulse types, including those with three-photon processes. 

\begin{acknowledgments}
The computing for this project was performed on the Beocat Research Cluster at 
Kansas State University, which is funded in part by NSF grants CNS-1006860, EPS-1006860, and EPS-0919443,
and used resources of the National Energy Research Scientific
Computing Center (NERSC), a U.S. Department of Energy Office of Science User
Facility operated under Contract No. DE-AC02-05CH11231.
CPK acknowledges financial support from the Deutsche Forschungsgemeinschaft (CRC
    1319). 
\end{acknowledgments}

\appendix
\section{Frame rotations}
\label{sec:appendix1}
For completeness, we provide in this section the details of the derivation of the
relevant quantities involving the frame rotations presented in the text. Hartree atomic units (a.u.) 
are used throughout.

In the laboratory frame $\mathcal{R}^\prime$, the electric field $\text{\textbf{E}}^\prime(t)$  
reads, 
\begin{subequations} 
\begin{eqnarray}
  \label{eq:Efield.def}
  \text{\textbf{E}}^\prime(t)=\sum_{\mu_0=0,\pm 1}\mathscr{E}^{\prime}_{\mu_0}(t)\,\text{\textbf{e}}^{*\prime}_{\mu_0},
\end{eqnarray}
where $(*)$ denotes the complex conjugation, and $\text{\textbf{e}}^{*\prime}_{\mu_0}$
  the spherical unit vectors such that 
\begin{eqnarray}
  \text{\textbf{E}}^\prime(t)\cdot\text{\textbf{e}}^\prime_{\mu_0} = \mathscr{E}^{\prime}_{\mu_0}(t). 
\end{eqnarray}
  As stated in the text,  $\mathscr{E}^\prime_{\mu_0}(t)$, for $\mu_0=\pm 1, 0$, refer to polarization unit components of the field in
  $\mathcal{R}^\prime$. In this frame, the spherical unit vectors are defined 
  in terms of their cartesian counterparts, 
  \begin{eqnarray}
    \label{eq:cart_to_sph}
    \text{\textbf{e}}^\prime_{\pm} = \mp 
    \dfrac{1}{\sqrt{2}}\big(\text{\textbf{e}}^\prime_{\text{x}}\pm
    \text{\textbf{e}}^\prime_{\text{y}}\big) 
  \end{eqnarray} 
  together with $\text{\textbf{e}}^\prime_0=\text{\textbf{e}}^\prime_z$. Projection 
  of the field components into the molecular frame --which is rotated of Euler
  angles~\cite{edmonds2016angular,rose1957elementary} $\gamma_{\mathcal{R}}$ with
respect to the laboratory frame-- is obtained by writing the spherical unit
  components $\text{\textbf{e}}^\prime_{\mu_0}$ in terms of their counterparts
  $\text{\textbf{e}}_{\mu_0}$ associated with the molecular frame,   
\begin{eqnarray}
  \label{eq:transformation.direct}
  \text{\textbf{e}}^\prime_{\mu_0} = \sum_{\mu=0,\pm 1} \mathcal{D}^{(1)}_{\mu, \mu_0}(\gamma_{\mathcal{R}})\,\text{\textbf{e}}_{\mu}\,. 
\end{eqnarray}
Recalling that $\text{\textbf{e}}^*_{\mu_0}=(-1)^{\mu_0}\text{\textbf{e}}_{-\mu_0}$, we find 
 \begin{eqnarray}
   \label{eq:efield_molframe}
   \text{\textbf{E}}(t;\gamma_{\mathcal{R}}) &=&\sum_{\mu,\mu_0}(-1)^{\mu_0}\mathscr{E}^{\prime}_{\mu_0}(t) \mathcal{D}^{(1)}_{\mu, \mu_0}(\gamma_{\mathcal{R}})\,\text{\textbf{e}}_{\mu}\,,
 \end{eqnarray}
 where $\mathcal{D}^{(1)}_{\mu,\mu_{0}}(\gamma_{\mathcal{R}})$ correspond the
elements of the Wigner rotation matrix~\cite{edmonds2016angular,rose1957elementary}. 
Similarly, we decompose the position operator into the spherical unit basis, namely 
\begin{eqnarray} 
  \Op{r}=\sum_{\mu^\prime=0,\pm 1}
  \Op{r}_{\mu^\prime}\text{\textbf{e}}^*_{\mu^\prime}\,.
\end{eqnarray}
   Using Eq.~\eqref{eq:efield_molframe}, the  molecular-frame orientation-dependent dipole interaction reads
\begin{eqnarray} 
  \text{\textbf{E}}(t;\gamma_{\mathcal{R}})\cdot\Op{r}=\sum_{\mu_0}(-1)^{\mu_0}\mathscr{E}^\prime_{\mu_0}(t)\sum_{\mu}\mathcal{D}^{(1)}_{\mu, -\mu_0}(\gamma_{\mathcal{R}})\,\Op{r}_{\mu}\,,\quad\quad\quad 
\end{eqnarray}
  which corresponds to the dipole interaction in the molecular frame $\mathcal{R}$ given in Eq.~\eqref{eq:dipol_mol_fram} in the text.
\end{subequations}

A second kind of rotation operations involves projection of the direction 
of photoelectron emission from the molecular to the laboratory frame coordinates as defined in
Eq.~\eqref{eq:epolyexpansion_labframe}. This results in the expressions for the first and second order
corrections in the laboratory frame of reference outlined in Eqs.~\eqref{eq:alpha_ik_first.final}
and~\eqref{eq:alpha_ik_sec.preliminary}, respectively. However, instead of
calculating $\alpha^{\boldsymbol{k^\prime}(n)}_{i_0}(t;\gamma_{\mathcal{R}})\cdot\alpha^{\boldsymbol{k^\prime}(m)*}_{i_0}(t;\gamma_{\mathcal{R}})$,
directly from Eqs.~\eqref{eq:alpha_ik_first.final} and~\eqref{eq:alpha_ik_sec.preliminary}, 
it turns out to be more convenient to rotate the 
anisotropy parameters themselves instead of calculating the anisotropy parameters
from the rotated expansion coefficients. Specifically, we follow  the 
prescription: 
\begin{enumerate}[label=(\alph*)] 
  \item Keep the ionization amplitudes $\alpha^{\boldsymbol{k}(n)}_{i_0}(t;\gamma_{\mathcal{R}})$, in
    the molecular frame, defined by  
    \begin{subequations}
    \begin{eqnarray}
  \label{eq:alpha_ik_first.final_molframe}
  \alpha^{\boldsymbol{k}(1)}_{i_0}(t;\gamma_{\mathcal{R}}) &=&
  i\sum_{\mu_0,\mu}(-1)^{\mu_0}\sum_{\ell,m}\mathcal{D}^{(1)}_{\mu,-\mu_0}(\gamma_{\mathcal{R}})\\
  &&
  \quad\times 
      (\mathbf{r}_{k,\ell,m;i_0}\cdot\textbf{e}_{\mu})\,
      Y^{\ell}_{m^\prime}(\theta_{\boldsymbol{k}},\phi_{\boldsymbol{k}}) 
      \nonumber\\
      &&\quad\times\int_{-\infty}^{t}e^{-i(\epsilon_0-\epsilon^k_{i_0})}\mathscr{E}^\prime_{\mu_0}(t^\prime)\,\mathrm{d}t^\prime\,.\nonumber 
\end{eqnarray}
for first order processes, and
      \begin{widetext}
      \begin{eqnarray}
  \label{eq:alpha_ik_sec.preliminary_molframe}
  \alpha^{\boldsymbol{k}(2)}_{i_0}(t;\gamma_{\mathcal{R}}) &=&
  -\sum_{\mu_0,\nu_0}(-1)^{\mu_0+\nu_0}\sum_{\mu,\nu}\mathcal{D}^{(1)}_{\mu,-\mu_0}(\gamma_{\mathcal{R}})\mathcal{D}^{(1)}_{\nu,-\nu_0}(\gamma_{\mathcal{R}})
        \sum_{\ell, m}\,
        Y^{\ell}_{m^\prime}(\theta_{\boldsymbol{k}},\phi_{\boldsymbol{k}})\nonumber\\
      &&\quad\times\Big[ \left(\mathbf{r}_{k,\ell,m;i_0}\cdot\textbf{e}_{\mu}\right) \sum_i \left(\mathbf{r}_{i,i}\cdot\textbf{e}_{\nu}\right) \,
      \int^t_{-\infty}e^{-i(\epsilon_{i_0}-\epsilon_k)t^\prime}
        \mathscr{E}^{\prime}_{\mu_0}(t^\prime)\,
      \int_{-\infty}^{t^\prime} \mathscr{E}^{\prime}_{\nu_0}(t^{\prime\prime})
        \,\mathrm{d}t^{\prime\prime}\,\mathrm{d}t^\prime\nonumber\\ %[0.3cm]
      &&\quad\quad +\sum_{b} \left(\mathbf{r}_{k,\ell,m;b}\cdot\textbf{e}_{\mu}\right)
      \left(\mathbf{r}_{b,i_0}\cdot\textbf{e}_{\nu}\right)
      \int^t_{-\infty}e^{-i(\epsilon_{b}-\epsilon_k)t^\prime}\mathscr{E}^{\prime}_{\mu_0}(t^\prime)\,
      \int_{-\infty}^{t^\prime}e^{-i(\epsilon_{i_0}-\epsilon_b)}\mathscr{E}^{\prime}_{\nu_0}(t^{\prime\prime})\,dt^{\prime\prime}\,dt^\prime\Big]\,.  % \\ %[0.3cm]
 \end{eqnarray}
\end{widetext}
for second order processes. For reasons that will become clearer later, we seek to express the products
$\alpha^{\boldsymbol{k}(n)}_{i_0}(t;\gamma_{\mathcal{R}})\alpha^{\boldsymbol{k}(m)*}_{i_0}(t;\gamma_{\mathcal{R}})$
in terms of a product involving three Wigner rotation matrices. To this end,
it is convenient to express the product $\mathcal{D}^{(1)}_{\mu,-\mu_0}(\gamma_{\mathcal{R}})\mathcal{D}^{(1)}_{\nu,-\nu_0}(\gamma_{\mathcal{R}})$ 
in Eq.~\eqref{eq:alpha_ik_sec.preliminary_molframe} in terms of its irreducible
representation using the elementary expression~\cite{edmonds2016angular,rose1957elementary} 
\begin{widetext}
\begin{eqnarray}
\label{subeq:property.product}
&&\mathcal{D}^{(\ell_1)}_{m_1,m^\prime_1}(\gamma_{\mathcal{R}})
\mathcal{D}^{(\ell_2)}_{m_2,m^\prime_2}(\gamma_{\mathcal{R}})
 =\sum_{\ell}
 (2\ell+1)\,\mathcal{D}^{*(j)}_{-m_{12},-m^\prime_{12}}(\gamma_{\mathcal{R}})
        \nonumber
       \begin{pmatrix} \ell_1 & \ell_2 & \ell \vspace*{0.33cm} \\ 
        m_1 & m_2 & -m_{12} 
       \end{pmatrix}
       \begin{pmatrix} \ell_1 & \ell_2 & \ell \vspace*{0.33cm} \\ 
       m^\prime_1 & m^\prime_2 & -m^\prime_{12} 
       \end{pmatrix}\,,
\end{eqnarray}
\end{widetext}
  with $m_{12}=m_1+m_2$ and $m^\prime_{12}=m^\prime_1+m^\prime_2$.
\end{subequations}
\item When calculating $\alpha^{\boldsymbol{k}(n)}_{i_0}(t;\gamma_{\mathcal{R}})\alpha^{\boldsymbol{k}(m)*}_{i_0}(t;\gamma_{\mathcal{R}})$
a product involving the spherical harmonics, $Y^{\ell}_{m^\prime}(\theta_{\boldsymbol{k}},\phi_{\boldsymbol{k}})Y^{\ell^\prime *}_{m^\prime}(\theta_{\boldsymbol{k}},\phi_{\boldsymbol{k}}) $ appears. The strategy is to first
express such  a product in terms of its irreducible representation, namely
\begin{eqnarray}
  \label{eq:mymethod}
  Y^{\ell}_{m}(\Omega_{\boldsymbol{k}})Y^{\ell^\prime*}_{m^\prime}(\Omega_{\boldsymbol{k}}) &=& (-1)^{m^\prime}
  \sum_{L=|\ell=\ell^\prime|}^{\ell+\ell^\prime} \eta^{L}_{\ell,\ell^\prime}\, Y^{L*}_{m^\prime-m}(\Omega_{\boldsymbol{k}})\quad\quad\,\,\nonumber\\
  &&
  \times\begin{pmatrix} \ell & \ell^\prime & L \vspace*{0.33cm} \\ 
  0 & 0 & 0 
  \end{pmatrix}
  \begin{pmatrix} \ell & \ell^\prime & L \vspace*{0.33cm} \\ 
  m & -m^\prime & m^\prime-m 
  \end{pmatrix}\,,\nonumber\\
\end{eqnarray}
with $\eta^{L}_{\ell,\ell^\prime}=\sqrt{(2\ell+1)(2\ell^\prime+1)/4\pi}$  
 and where we have defined $\Omega_{\boldsymbol{k}^\prime} = (\theta_{\boldsymbol{k}^\prime},\phi_{\boldsymbol{k}^\prime})$.
\item
 Next, we rotate 
  the resulting spherical harmonic in Eq.~\eqref{eq:mymethod} in the
  laboratory frame, $\mathcal{R}^\prime$, using the inverse of the frame transformation
  defined in Eq.~\eqref{eq:transformation.direct}, namely, 
  \begin{eqnarray}
    \label{eq:MyY}
    Y^{L}_{m-m^\prime}(\Omega_{\boldsymbol{k}}) = \sum^L_{M=-L}
    \mathcal{D}^{(1)\dagger}_{M,m-m^\prime}(\gamma_{\mathcal{R}})Y^{L}_{M}(\Omega_{\boldsymbol{k}^\prime})
  \end{eqnarray}

\item Finally, we write $Y^L_M(\Omega_{\boldsymbol{k}^\prime})$ in Eq.~\eqref{eq:MyY} in terms of the
  associated Legendre polynomials~\cite{edmonds2016angular,rose1957elementary}
 \begin{eqnarray}
   \label{eq:Yfinal}
   Y^{L}_{M}(\Omega_{\boldsymbol{k}^\prime}) &=& (-1)^M\,\sqrt{\dfrac{(2L+1)}{4\pi}\dfrac{(L-M)!}{(L+M)!}}\nonumber\\
   &&\times P^M_L(\cos\theta_{\boldsymbol{k}^\prime})\, e^{iM\phi_{\boldsymbol{k}^\prime}}\,, 
 \end{eqnarray}
 \item following these steps, Eq.~\eqref{eq:mymethod} finally reads,
\begin{widetext}
\begin{eqnarray}
  Y^{\ell}_{m^\prime}(\theta_{\boldsymbol{k}},\phi_{\boldsymbol{k}})Y^{\ell^\prime *}_{m^\prime}(\theta_{\boldsymbol{k}},\phi_{\boldsymbol{k}}) &=&
  (-1)^{m^\prime}\,\sum^{\ell+\ell^\prime}_{L=|\ell-\ell^\prime|}\dfrac{2L+1}{4\pi}\sqrt{(2\ell+1)(2\ell^\prime+1)}\, 
  \begin{pmatrix} \ell & \ell^\prime & L \vspace*{0.33cm} \\ 
  0 & 0 & 0 
  \end{pmatrix}
  \begin{pmatrix} \ell & \ell^\prime & L \vspace*{0.33cm} \\ 
  m & -m^\prime & m^\prime-m 
  \end{pmatrix} \nonumber\\
  && \times \sum^L_{M=-L}\sqrt{\dfrac{(L-M)!}{(L+M)!}}\, \mathcal{D}^{(L)}_{m^\prime-m,-M}(\gamma_{\mathcal{R}})\, 
 P^M_L(\cos\theta_{\boldsymbol{k}^\prime})e^{iM\phi_{\boldsymbol{k}^\prime}}\,, 
\end{eqnarray}
\end{widetext}
\end{enumerate}
which transforms the anisotropy parameters from the molecular to the laboratory
frame of reference. Thus, this strategy is equivalent to rotate the anisotropy parameters instead of
performing the full derivation using the rotated wave function coefficients.
Apart from significantly reducing the number of
operations in the summations when calculating the norm squared of first and
second order corrections, it has the advantage of leading to an expression for $M=-L,L$ in $\beta^{(\cdot)}_{L,M}(\epsilon_k)$, as an function of 
the polarization unit vectors $\epsilon_{\mu_0},\epsilon_{\nu_0}$ and $\epsilon_{\mu^\prime_0},\epsilon_{\nu^\prime_0}$ in a
straightforward manner, which facilitates the analysis of the selection rules
for the anisotropy as a function of the field polarization direction.

\section{Orientation-averaged anisotropy parameters}
\label{sec:appendix2}

\subsection{Anisotropy parameters $\beta^{1ph}_{L,M}$}
\label{subsubsec:beta11}
 Following the guidelines for rotating the anisotropy parameters introduced 
 in Sec.~\ref{sec:appendix1}, the orientation-averaged ionization probability distribution 
 for
 one-photon ionization
 measured in the laboratory frame, $\mathcal{R}^\prime$,  is obtained upon
 rotation of $|\alpha^{\boldsymbol{k}(1)}_{i_0}(t;\gamma_{\mathcal{R}})|^2$ --using Eq.~\eqref{eq:alpha_ik_first.final_molframe}--
 from the molecular to the laboratory frame  or reference,  and
 integrating over all Euler angles
 $\gamma_{\mathcal{R}}$. We find, 
  \begin{widetext}
\begin{eqnarray}
  \label{eq:PES1} \left|\alpha^{\boldsymbol{k}^\prime(1)}_{i_0}(\gamma_{\mathcal{R}})\right|^2 &=&\sum_{L, M}\dfrac{(2L+1)}{4\pi}\sqrt{\dfrac{(L-M)!}{(L+M)!}}
    \displaystyle\sum_{\mu_0}(-1)^{-\mu_0}\mathcal{I}_{\mu_0}(k,t)\sum_{\mu^\prime_0}\mathcal{I}^*_{\mu^\prime_0}(k,t)\sum_{\ell,m,\mu} \left(\mathbf{r}_{k,\ell,m;i_0}\cdot\textbf{e}_{\mu}\right)\sum_{\ell^\prime,m^\prime,
    \mu^\prime}\left(\mathbf{r}^*_{k,\ell^\prime,m^\prime;i_0}\cdot\textbf{e}_{\mu^\prime}\right)\nonumber\\
    &&\times(-1)^{\mu^\prime+m^\prime}
  \sqrt{(2\ell+1)(2\ell^\prime+1)}
 \begin{pmatrix} \ell & \ell^\prime & L\vspace*{0.31cm}\\ 0 & 0 & 0 \end{pmatrix} 
 \begin{pmatrix} \ell & \ell^\prime & L\vspace*{0.31cm}\\ m & -m^\prime &m^\prime-m\end{pmatrix}  
   P^{M}_{L}(\cos\theta_{\boldsymbol{k}^\prime})\,e^{+iM\phi_{\boldsymbol{k}^\prime}}
    \nonumber\\ &&\times
   \int\dfrac{d^3\gamma_{\mathcal{R}}}{8\pi^2}\mathcal{D}^{(1)}_{\mu,-\mu_0}(\gamma_{\mathcal{R}})
\mathcal{D}^{(1)}_{-\mu^\prime,\mu^\prime_0}(\gamma_{\mathcal{R}})
\mathcal{D}^{(L)}_{m^\prime-m, -M}(\gamma_{\mathcal{R}})
\end{eqnarray}
\end{widetext}
for $t\rightarrow\infty$ and where we have defined
%% \begin{eqnarray}
%%   \mathcal{M}^{\mu}_{k,\ell, m} =\langle\varphi^{-}_{k,\ell,
%%   m}|\Op{r}_{\mu}|\varphi_{i_0}\rangle 
%% \end{eqnarray}
%% and
\begin{eqnarray}
  \label{eq:Def.I}
  \mathcal{I}_{\mu_0}(k,t)
  =\int^{t}_{-\infty}\mathscr{E}^\prime_{\mu_0}(t^\prime)\,e^{i(\epsilon_k-\epsilon_{i_0})t^\prime}\,dt^\prime \,.
\end{eqnarray}
Integration of Eq.~\eqref{eq:PES1} over the Euler angles  defines, according to
Eq.~\eqref{eq:1phcontr}, the laboratory-frame orientation-averaged anisotropy parameter $\beta^{(\mu_0)1ph}_{L,M}(\epsilon_k)$
corresponding to the first order process. Integration of a product involving three Wigner $3j-$ symbols 
can be performed analytically,~\cite{edmonds2016angular,rose1957elementary}
\begin{eqnarray}
  \label{eq:3j.int}
&& \int\mathcal{D}^{(\ell_1)}_{m_1,m^\prime_1}(\gamma_{\mathcal{R}}) \mathcal{D}^{(\ell_2)}_{m_2,m^\prime_2}(\gamma_{\mathcal{R}})\mathcal{D}^{(\ell_3)}_{m_3,m^\prime_3}(\gamma_{\mathcal{R}}) \dfrac{d^3\gamma_{\mathcal{R}}}{8\pi^2}\nonumber\\[0.2cm] 
         &&\quad\quad\quad=\begin{pmatrix} \ell_1 & \ell_2 & \ell_3\vspace*{0.31cm}\\ m_1 & m_2 & m_3\end{pmatrix} 
\begin{pmatrix} \ell_1 & \ell_2 & \ell_3\vspace*{0.31cm}\\ m^\prime_1 & m^\prime_2 & m^\prime_3\end{pmatrix}\,.\\[0.2cm]\nonumber 
\end{eqnarray}
which gives, upon equating Eq.~\eqref{eq:1phcontr} and Eq.~\eqref{eq:PES1}, the final expression,
\begin{widetext}
\begin{eqnarray}
  \label{eq:beta1.final_appendix} \beta^{1ph}_{L,M}(\epsilon_k)&=&\dfrac{(2L+1)}{4\pi}\sqrt{\dfrac{(L-M)!}{(L+M)!}}
    \sum_{\mu_0}(-1)^{-\mu_0}\mathcal{I}_{\mu_0}(k,t)\sum_{\mu^\prime_0}\mathcal{I}^*_{\mu^\prime_0}(k,t)\sum_{\ell,m,\mu}\left(\mathbf{r}_{k,\ell,m;i_0}\cdot\textbf{e}_{\mu}\right)\sum_{\ell^\prime,m^\prime,
    \mu^\prime}\left(\mathbf{r}_{k,\ell^\prime,m^\prime;i_0}\cdot\textbf{e}_{\mu^\prime}\right)^{*}\\
  &&\times(-1)^{\mu^\prime+m^\prime}\, \sqrt{(2\ell+1)(2\ell^\prime+1)}\times
 \begin{pmatrix} \ell & \ell^\prime & L\vspace*{0.31cm}\\ 0 & 0 & 0 \end{pmatrix} 
 \begin{pmatrix} \ell & \ell^\prime & L\vspace*{0.31cm}\\ m & -m^\prime &m^\prime-m\end{pmatrix}  
 \begin{pmatrix} 1 & 1 & L\vspace*{0.31cm}\\ \mu & -\mu^\prime &m^\prime-m\end{pmatrix}  
 \begin{pmatrix} 1 & 1 & L\vspace*{0.31cm}\\ -\mu_0 & \mu^\prime_0 &-M\end{pmatrix}\,,\quad\quad\nonumber  
\end{eqnarray}
\end{widetext}
which corresponds to the quantity displayed in Eq.~\eqref{eq:beta1.final} in the text.

\subsection{Anisotropy parameters $\beta^{2ph}_{L,M}$}
\label{subsubsec:beta22}
To evaluate the laboratory-frame orientation-averaged anisotropy parameters describing
the contribution from second order processes, $\beta^{2ph}_{L,M}(\epsilon_k)$,
we employ the same stratery involving the elementary angular momentum algebra detailed in Sec.~\ref{subsubsec:beta11}.  
Using Eq.~\eqref{eq:alpha_ik_sec.preliminary_molframe} and upon evaluation of
the product $\alpha^{\boldsymbol{k}(2)}_{\i_0}(\epsilon_k)\cdot\alpha^{\boldsymbol{k}(2)*}_{\i_0}(\epsilon_k)$, followed by projection of quantity into
the laboratory frame coordinates as indicated above, we find, with the help of Eq.~\eqref{eq:1phcontr}, 
\begin{widetext}
\begin{eqnarray}
  \label{eq:beta2.final_appendix} 
  \beta^{2ph}_{L,M}(\epsilon_k)&=&\dfrac{(2L+1)}{4\pi}\sqrt{\dfrac{(L-M)!}{(L+M)!}}
  \sum_{\mu_0,\nu_0}\sum_{\mu,\nu}(-1)^{-\mu-\nu}\sum_{Q_1=0}^2 g^{(Q_1)}_{\mu,\nu,\mu_0,\nu_0}
  \sum_{\mu^\prime_0,\nu^\prime_0}\sum_{\mu^\prime,\nu^\prime}(-1)^{\mu^\prime_0+\nu^\prime_0}\sum_{Q_2=0}^2 g^{(Q_2)}_{\mu^\prime,\nu^\prime,\mu^\prime_0,\nu^\prime_0}\nonumber\\
  &&\times\sum_{\ell, m}\sum_{\ell^\prime, m^\prime}(-1)^{m^\prime}\sqrt{(2\ell+1)(2\ell^\prime+1)}\sum_{p,p^\prime\ge i_0}
  \mathcal{S}^{p^\prime*}_{\mu^\prime,\nu^\prime}(k,\ell^\prime,m^\prime)\,\mathcal{F}^{p^\prime *}_{\mu^\prime_0,\nu^\prime_0}(t;k)\,
  \mathcal{S}^{p}_{\mu,\nu}(k,\ell,m)\,\mathcal{F}^{p}_{\mu_0,\nu_0}(t;k)\,
  \nonumber\\
  && \times
  \begin{pmatrix} \ell & \ell^\prime & L\vspace*{0.31cm}\\ 0 & 0 & 0\end{pmatrix}  
  \begin{pmatrix} \ell & \ell^\prime & L\vspace*{0.31cm}\\ m & -m^\prime &m^\prime-m\end{pmatrix}  
  \begin{pmatrix} Q_1 & Q_2 & L\vspace*{0.31cm}\\ \mu+\nu & -\mu^\prime-\nu^\prime &m^\prime-m\end{pmatrix}  
    \begin{pmatrix} Q_1 & Q_2 & L\vspace*{0.31cm}\\ -\mu_0-\nu_0 & \mu^\prime_0+\nu^\prime_0& -M\end{pmatrix}\,. 
\end{eqnarray}
\end{widetext}
Note that the limits on the sum over $Q_1$ and $Q_2$ imply
that the two-photon processes contribute with a polynomial order of $L=4$ 
at the most. In Eq.~\eqref{eq:beta2.final} we have defined,
\begin{subequations}
\begin{eqnarray}
  \label{eq.gQ.def}
  g^{(Q)}_{\mu,\mu^\prime, \mu^\prime_0,\nu^\prime_0} &\equiv& c_Q\,
  \begin{pmatrix} 1 & 1 & Q\vspace*{0.31cm}\\ \mu & \nu& -\mu-\nu \end{pmatrix}
    \begin{pmatrix} 1 & 1 & Q\vspace*{0.31cm}\\ -\mu_0 & -\nu_0 & \mu_0+\nu_0 \end{pmatrix}\,.\nonumber\\
\end{eqnarray}
 %with $c_Q = (2Q+1)$. Next, using 
 %$\mathbf{r}_{k,\ell,m;p}=\langle\varphi^{-}_{k\ell,m}|\op{r}|\varphi_p\rangle$ and
 %$\mathbf{r}_{p,i}=\langle\varphi_p|\op{r}|\varphi_{i}\rangle$ as defined in the text, the quantity $\mathcal{S}^{p}_{\mu,\nu}(k,\ell,m)$
 %in Eqs.~\eqref{eq:beta2.final_appendix} and~\eqref{eq:beta2.final} is given by,  
  with $c_Q = (2Q+1)$. The term  $\mathcal{F}^{p}_{\mu_0,\nu_0}(t;k)$ in Eq.~\eqref{eq:beta2.final_appendix} is given by, 
  \begin{eqnarray}
  \label{eq.F.def}
    \mathcal{F}^{p}_{\mu_0,\nu_0}(t;k)&=&
  \int^t_{-\infty}\!\!\!\!\!\!\!e^{i(\epsilon_k-\epsilon_p)t^\prime}\mathscr{E}^\prime_{\mu_0}(t^\prime)\nonumber\\
    &&\quad\quad\quad\times\int_{-\infty}^{t^\prime}\!\!\!\!\!\!\!e^{i(\epsilon_p-\epsilon_{i_0})t^{\prime\prime}}\mathscr{E}^\prime_{\nu_0}(t^{\prime\prime})dt^\prime dt^{\prime\prime}\,.\quad\nonumber\\ 
\end{eqnarray}
 Finally, the term $\mathcal{S}^{p}_{\mu,\nu}(k,\ell,m)$ in Eq.~\eqref{eq:beta2.final_appendix} reads 
  \begin{eqnarray}
  \label{eq.Sa.def}
  \mathcal{S}^{p}_{\mu,\nu}(k,\ell,m) &=&
    (1-\delta_{p,i_0}) \left(\mathbf{r}_{k,\ell,m;p}\cdot\textbf{e}_{\mu} \right)\, \left(\mathbf{r}_{p,i_0}\cdot\textbf{e}_{\nu} \right)\nonumber\\
    &&+\delta_{p,i_0}\sum_{i\in occ} \left(\mathbf{r}_{k,\ell,m;p}\cdot\textbf{e}_{\mu}\right)\,\left(\mathbf{r}_{i;i}\cdot\textbf{e}_{\nu}\right)\,,\nonumber\\
\end{eqnarray}
  for $p\ge i_0$. 
\end{subequations}
From Eq.~\eqref{eq.gQ.def}
and the fourth Wigner $3j-$symbol in Eq.~\eqref{eq:beta2.final_appendix}, it follows that second-order processes also lead
to vanishing asymmetries in the PAD for linearly polarized fields.  
In fact,  Eq.~\eqref{eq.gQ.def} vanishes for  $Q$ odd, while the fourth Wigner $3j-$symbol Eq.~\eqref{eq:beta2.final} requires $Q_1+Q_2$ to be odd for $L$ odd.

\subsection{Anisotropy parameters $\beta^{int}_{L,M}$}
\label{subsubsec:beta12}

The contribution to the photoelectron momentum distribution originating from the interference between 
single- and two-photon pathways are obtained using the same strategy employed in 
Secs.~\ref{subsubsec:beta11} and~\ref{subsubsec:beta22}. Upon straightforward
angular momentum algebra,  we find 
\begin{widetext}
 \begin{eqnarray}
 \label{eq:beta12.final_appendix} 
   \beta^{int}_{L,M}(\epsilon_k)&=&-i\dfrac{(2L+1)}{4\pi}\sqrt{\dfrac{(L-M)!}{(L+M)!}}
   \sum_{\mu,\ell,m}\left(\mathbf{r}_{k,\ell,m;i_0}\cdot\textbf{e}_{\mu}\right)\sum_{\mu_0}(-1)^{-\mu_0}\int^{t}_{-\infty}\mathscr{E}^\prime_{\mu_0}(t^\prime)\,e^{+(\epsilon_k-\epsilon_{i_0})t^\prime}\,dt^\prime
   \\
   &&\times\sum_{\mu^\prime_0, \nu^\prime_0}(-1)^{\mu^\prime_0+\nu^\prime_0}\sum_{\mu^\prime,\nu^\prime}\sum^{2}_{Q2=0}g^{(Q_2)}_{\mu^\prime, \nu^\prime, \mu^\prime_0, \nu^\prime_0} \sum_{\ell^\prime, m^\prime}(-1)^{m^\prime}\,\sum_{p\ge i_0} 
  \mathcal{S}^{p^\prime*}_{\mu^\prime,\nu^\prime}(k,\ell^\prime,m^\prime)\,
  \mathcal{F}^{p^\prime *}_{\mu^\prime_0,\nu^\prime_0}(t;k)\,
  \nonumber\\
  &&\times\sqrt{(2\ell+1)(2\ell^\prime+1)}\, 
 \begin{pmatrix} \ell & \ell^\prime & L \vspace*{0.31cm}\\ 0 & 0 & 0 \end{pmatrix} 
 \begin{pmatrix} \ell & \ell^\prime & L \vspace*{0.31cm}\\ m & -m^\prime & m^\prime-m \end{pmatrix} 
 \begin{pmatrix} 1 & Q_2 & L \vspace*{0.31cm}\\ \mu & -\mu^\prime-\nu^\prime & m^\prime-m \end{pmatrix} 
 \begin{pmatrix} 1 & Q_2 & L \vspace*{0.31cm}\\ -\mu_0 & \mu^\prime_0+\nu^\prime_0 & -M \end{pmatrix}\,,\nonumber 
 \end{eqnarray}
 \end{widetext}
for $t\rightarrow\infty$ and where the terms $g^{(Q_2)}_{\mu,\nu,\mu_0,\nu_0}$, $\mathcal{S}^{p^\prime*}_{\mu^\prime,\nu^\prime}(k,\ell^\prime,m^\prime)$
and $\mathcal{F}^{p^\prime *}_{\mu^\prime_0,\nu^\prime_0}(t;k)$ 
in Eq.~\eqref{eq:beta12.final_appendix} are given in Eq.~\eqref{eq.gQ.def}, ~\eqref{eq.Sa.def} and~\eqref{eq.F.def}, respectively.

From the third and fourth Wigner $3j$-symbols in Eq.~\eqref{eq:beta12.final_appendix}, 
it is apparent that the interference term, $\beta^{int}_{L,M}$ may contribute to the
anisotropy after the orientation-averaging, even for linearly polarized fields. In fact,
for $L$ odd and $Q_2$ even, both symbols do not necessarily vanish by selection
rules, in contrast to $\beta^{1ph}_{L,M}$ and $\beta^{2ph}_{L,M}$. Conversely, for circularly polarized fields, or fields with
unequal counter-rotating components, all three orientation-averaged anisotropy
parameters may contribute to the
anisotropy in the photoelectron emission in the case of a chiral target. For
achiral targets, only interference term can be used to break the asymmetry.

\section{Pulse parametrization for polarization shaped pulses}
\label{sec:polarization_param}

For polarization shaped pulses, we consider a superposition of
pulses with different counter-rotating components, namely 
\begin{subequations}
  \label{eq:combinationRL}
  \begin{eqnarray}
    \text{\textbf{E}}^{\prime}(t) =\text{\textbf{E}}^{\prime}_{\text{R}}(t)+\text{\textbf{E}}^{\prime}_{\text{L}}(t). 
\end{eqnarray}
  Each rotating component, carrying
  circularly left and right polarization, $\text{\textbf{E}}^{\prime}_{\text{L,R}}(t)$, is projected into the
  polarization unit vectors, $\text{\textbf{e}}^{\prime}_{\pm}$, according to,
\begin{eqnarray}
  \text{\textbf{E}}^{\prime}_{\text{L,R}}(t)
  = \mathscr{E}^{\prime \text{L,R}}_{+}(t)\, \text{\textbf{e}}^{\prime *}_{+1} + \mathscr{E}^{\prime \text{L,R}}_{-}\, \text{\textbf{e}}^{\prime *}_{-1}\,,
\end{eqnarray}
  where the CRP and CLP components $\mathscr{E}^{\prime\,\text{L,R}}_{\pm}=\text{\textbf{E}}^\prime_{\text{L,R}}\cdot\text{\textbf{e}}^{\prime}_{\pm}$ are (independently) 
  parametrized according to, 
  \begin{eqnarray}
   \label{eq:pulseparamR}
  \mathscr{E}^{\prime R}_{\pm}(t) = \pm\dfrac{1}{\sqrt{2}}\sum^N_{j=1} h_{j}(t-\tau_j)\,e^{\pm i\omega_j(t-\tau_j)+\phi_j}\,
\end{eqnarray}
  for circularly right polarization (CRP). Its counter-rotating couterpart takes the form 
  \begin{eqnarray}
   \label{eq:pulseparamL}
    \mathscr{E}^{\prime L}_{\pm}(t) = \mp\dfrac{1}{\sqrt{2}}\sum^N_{j=1} h_{j}(t-\tau_j)\,e^{\mp i\omega_j(t-\tau_j)+\phi_j}\,
\end{eqnarray}
which allows to retrieve the rotating field components in cartesian coordinates  
\begin{eqnarray}
  \text{\textbf{E}}^{\prime}_{\text{L,R}} =
  \sum^N_{j=1}h(t-\tau_j) 
  \begin{pmatrix} 
    \pm \cos\Omega_j(t) \\[0.2cm] 
    -\sin\Omega_j(t)\\[0.2cm]
    0
  \end{pmatrix}
  \end{eqnarray}
\end{subequations}
for circularly right $(-)$ and left $(+)$ polarization directions from the
source point of view in the laboratory frame of reference and where $\Omega_j(t) =\omega_j(t-\tau_j) + \phi_j $. 
The pulse parameters in Eqs.~\eqref{eq:pulseparamR} and~\eqref{eq:pulseparamL} 
are independently optimized for both counter-rotating components.

\end{document}